\documentclass[preprint]{aastex}
\usepackage{graphics}
\usepackage{amssymb}
\usepackage{amsmath}

\slugcomment{Published in AJ}
\shorttitle{Selection
effect of radio radio quasars in SDSS} \shortauthors{Lu et al.}

\begin{document}
\title{On the selection effect of radio quasars in the Sloan Digital Sky Survey}
\author{Yu Lu\altaffilmark{1,2}, Tinggui Wang\altaffilmark{1,2}, Hongyan Zhou\altaffilmark{1,2},
Jian Wu\altaffilmark{3}}
\altaffiltext{1}{Center for Astrophysics, University of Science and Technology
       of China, Hefei, Anhui, 230026, P.R.China}
\altaffiltext{2}{Joint Institute of Galaxies and Cosmology, SHAO and USTC,
                 Hefei, Anhui, 230026, China}
\altaffiltext{3}{Department of Astronomy \& Astrophysics, the Pennsylvania
                 State University,}
\email{laolu@mail.ustc.edu.cn}

\begin{abstract}
We identified a large sample of radio quasars, including those with
complex radio morphology, from the Sloan Digital Sky Survey (SDSS)
and the Faint Images of Radio Sky at Twenty-cm (FIRST). Using this
sample, we inspect previous radio quasar samples for selection
effects resulting from complex radio morphologies and adopting
positional coincidence between radio and optical sources alone. We
find that $13.0$\% and $8.1$\% radio quasars do not show a radio
core within $1 \farcs 2$ and $2^{''}$ of their optical position, and
thus are missed in such samples. Radio flux is under-estimated by a
factor of more than 2 for an additional $8.7\%$ radio quasars. These
missing radio extended quasars are more radio loud with a typical
radio-to-optical flux ratio namely radio loudness $RL\gtrsim 100$,
and radio power $P\gtrsim 10^{25}$ W Hz$^{-1}$. They account for
more than one third of all quasars with $RL>100$. The color of radio
extended quasars tends to be bluer than the radio compact quasars.
This suggests that radio extended quasars are more radio powerful
sources, e.g., Fanaroff-Riley type 2 (FR-II) sources, rather than
the compact ones viewed at larger inclination angles. By comparison
with the radio data from the NRAO VLA Sky Survey (NVSS), we find
that for sources with total radio flux less than 3 mJy, low surface
brightness components tend to be underestimated by FIRST, indicating
that lobes in these faint radio sources are still missed.
\end{abstract}

\section{Introduction}

In past decades, we have witnessed a rapid growth in the number of
radio selected Active Galactic Nuclei (AGNs) resulting from large
and deep radio surveys such as the Faint Images of Radio Sky at
Twenty centimeters (FIRST, Becker, White \& Helfand 1995) and the
NRAO VLA Sky Survey (NVSS, Condon et al. 1998) coupled with optical
spectroscopy follow-ups or dedicated spectroscopic surveys such as
the Two degree Field (2dF, Maddox et al. 1998; Boyle et al. 2001)
and the Sloan Digital Sky Survey (SDSS, York et al. 2000; Stoughton
et al. 2002). However, some fundamental issues regarding the origin
of radio emission remain hotly debated: is the radio loudness
dichotomy true or not (Kellermann et al. 1989; Miller, Peacock \&
Mead 1990; Hewett et al. 2001; Ivezic et al. 2002)?  Are radio jets
in radio quiet/intermediate quasars relativistic (Readhead et al.
1988; Wilson \& Colbert 1995; Meier et al.  2001)? Which physical
parameters control the large range of radio strength despite their
great similarity in the SED at other wavelengths for various quasars
(Barthel 1989; Urry \& Padovani 1995; Jackson \& Wall 1999; Boroson
2002; Aars et al. 2005)?

Concerning the first question, the ambiguity is caused largely by
various selection biases introduced by the survey limits, the
incompleteness in the radio sample due to their complex morphologies
or in the optical sample due to various color selections (Ivezic et
al. 2002; Cirasuolo et al. 2003a; Best et al. 2005). The traditional
radio surveys were carried out at shallow flux density limits (0.1-1
Jy) and primarily radio-loud quasars were detected (e.g., Bennett
1962; Smith \& Spinrad 1976; Colla et al. 1972; Fanti et al. 1974;
Large et al. 1981). Only the two latest radio surveys NVSS and FIRST
have enough sensitivity and positional accuracy to allow for the
detection of large number of radio intermediate and radio quiet
quasars in conjunction with moderately deep, large area optical
surveys such as 2dF and SDSS. However, by taking advantage of the
accuracy in the position of radio sources, most authors constructed
the radio quasar or quasar candidate sample using solely the
position match between radio and optical sources (Gregg et al. 1996;
White 1999; Lacy \& Ridgway 2001; McMahon et al  2002; Richards
et.al. 2002; Cirasuolo et al.2003b). This process introduces a bias
against lobe-dominated radio quasars: either they were missed or
their radio flux underestimated.

White, Becker \& Gregg (1999, 2000) argued that this incompleteness
is not severe in FIRST Bright Quasar Survey (FBQS), which was
selected by matching of optical counterparts within a $1\farcs2$
position offset to the radio sources in the FIRST catalog. Using a
matching radius of $2^{''}$, Ivezic et al. (2002) estimated that
less than $10\%$ SDSS-FIRST associations have complex radio
morphology, and core-lobe and double-lobe sources together represent
about only $5\%$ in the radio quasars and galaxy sample. Using a
novel technique, de Vries et al. (2006) constructed a Fanaroff-Riley
type 2 (FR-II) quasar sample, and found that $27\%$ FR-II quasars do
not show cores at the FIRST flux limit. These authors also compared
the emission line properties and optical colors of these FR II
quasars with radio quiet quasars. It should be noted that these
missing quasars are not random, but are all extended sources and
tend to be more radio loud (Falcke et al 1996; Ivezic et al. 2002;
Best et al 2005). As a result, the statistical properties of the
sample, such as radio loudness and radio luminosity distribution,
will be affected by this selection effect.

In this paper we study in detail of selection effects in the SDSS
radio quasar samples. We identified from SDSS and FIRST a large
sample of radio quasars, including those with complex radio
morphology. Besides using positional coincidence as a primary
selection criterion, we manually examined the FIRST images for all
of the candidates with extended radio morphology. Through this less
efficient process, we obtained a sample of $3641$ spectroscopically
confirmed quasars with secure radio identification. A detail
comparison of this sample with other radio quasars sample is given.
Various selection effects are quantified. Throughout this paper, we
will adopt a concordant cosmology with $H_0 = 70$
km~s$^{-1}$~Mpc$^{-1}$, $\Omega_m = 0.3$, and $\Omega_{\Lambda}=
0.7$.

\section{The Radio Quasar Sample}

\subsection{Optical data}

Our starting point is the SDSS quasar catalog constructed by Schneider
et al (2005). The catalog consists of quasars that contain at
 least one broad emission line ($FWHM \gtrsim 1000$ km~s$^{-1}$
), or are unambiguously broad absorption line quasars. These quasars
selected from SDSS photometric catalog either by their colors or for
their positional coincidence with radio sources in the FIRST catalog
(within 2\texttt{"} of a FIRST source) or ROSAT X-ray sources
(within 10-20\texttt{"} of a ROSAT source), were spectroscopically
confirmed to meet above criteria and their absolute optical
magnitudes at $i$-band $M_i\le -22.0$. The sample also includes some
supplementary quasars that meet the above criteria, but were
selected initially as galaxy targets (Eisenstein et al. 2001;
Strauss et al. 2002). Note that the magnitude limits for various
candidates are different, $i<19.1$ for color selected low $z$
($z<3.0$) quasars, $i<20.2$ for color-selected high $z$ quasars,
$i<19.1$ for FIRST and ROSAT sources, and $i<17.7$ for the main
galaxy sample. The final sample contains $46,420$ quasars in the
redshift range $0.078 \le z \le 5.414$, absolute magnitude range
$-30.2 \le M_{i} \le -22.0$, and $i$-band optical magnitude range
$15.10\le i \le 21.78$.

We use a subsample of this catalog in order to minimize the bias
introduced in the selection of quasar candidates. The SDSS sources
with their target-flags as ``QSO\_FIRST\_CAP'' or
``QSO\_FIRST\_SKIRT'' only (hereafter FIRST-only sources) will bias
against the lobe-dominated radio sources, and are redder than the
color-selected quasars (Richards et al. 2002). Therefore, they will
be excluded from the sample. Since radio and X-ray emission from
quasars are well-correlated (Shastri et al. 1993, Brinkmann et al.
2000; Padovani et al. 2003), the ROSAT-only selected quasars will be
biased toward radio strong sources, and as such are excluded from
the sub-sample.

\subsection{Radio data}
Radio counterparts to SDSS quasars are found using the FIRST survey.
The survey covers about 10,000 $deg^2$ and is $95$\% complete to 2
mJy, $80$\% complete to 1mJy (Becker et al 1995). The source surface
density in this survey is $\sim90$ deg$^{-2}$. At the 1 mJy source
detection threshold, the resolution is better than $5^{''}$.
Individual sources down to 1 mJy threshold have $90$\% confidence
error circles with radii of $\le 1^{''}$.

The FIRST catalog was produced by fitting a two-dimensional Gaussian
to each source to generate major axes, minor axes, peak and
integrated flux densities from the co-added images. The major axes
have been deconvolved to remove blurring by the elliptical Guassian
fitting. For bright sources, the size was determined down to $\sim
1/3$ of the beam size $5\farcs4$. The FIRST survey also provides
clean radio images.

\subsection{Compact radio quasars}
Most radio quasars are compact sources in the FIRST images and as
such, they can be identified through cross-correlation of the FIRST
catalog with the SDSS quasar catalog by adopting a small matching
radius in the position offset.

The matching radius is a trade-off between the completeness and
random association, i.e., the higher completeness necessarily
implies higher random contamination. Knapp et al. (2002) showed that
the random association increases with matching radius at $\leqslant
2\farcs5$. Magliocchetti \& Maddox (2002) found that $2^{''}$
matching radius could include $\sim97\%$ of the true matches above 1
mJy level. Gregg \& Becker (1996) estimated that more than $95\%$
FBQS-I quasars with magnitude $E\leqslant 17.8$ are within
$1\farcs1$ offset between the POSS-I and FIRST positions, while
White, Becker et al. (1999) showed that $\sim 1^{''}-1\farcs1$
matching radius will eliminate most of false quasar candidates at
the expense of only $5\%$ of incompleteness, and the fraction of
optical candidates found to be quasars declines steadily from $80\%$
near $0^{''}$ offset to $\sim 20\%$ at $1\farcs2$, and is constant
farther out. Ivezic et al. (2002) estimated that $1^{''}$ matching
radius produces $72\%$ completeness and $1.5\%$ random
contamination, and $1\farcs5$ matching radius with $85\%$
completeness and a contamination of 3\%. For their photometric SDSS
quasar sample, they estimate $3^{''}$ matching radius will cover
almost all counterpart but with 9\% random contamination.

Following Richards et al. (2002), we used a matching radius of
$2^{''}$ for compact FIRST sources and obtained 2782 matches with a
false rate\footnote{The fraction of chance coincidence is estimated
as $\rho \pi r^{2} N_{s}/N_{m}$, where $\rho $ is the surface
density of FIRST sources, $r$ the matching radius, $N_{s}$ the
number of quasars in the catalog of SDSS DR3, and $N_{m}$ the number
of matches.} of $\sim 0.15\%$. In addition, we found 71 SDSS quasars
locate within the ellipses of FIRST sources with core-jet or diffuse
structure, but with their optical-radio offset larger than the
$2^{''}$ matching radius. We added these quasars to our sample after
visual confirmation of the true association.

\subsection{Extended radio quasars}

The two point angular correlation function (Cress et al.  1995) can
be used to define an appropriate scale for searching for radio
matches with complex radio morphologies. The correlation function in
$0.02-2^{\circ}$ can be well fitted by a power law $A \times
{\theta}^{\gamma}$, where the angle $\theta$ in units of degree,
$A\sim3 \times 10^{-3}$ and ${\gamma}\sim-1.1$. At the ${\theta}
\sim 0.1^{\circ}$ ($6^{\prime}$), it drops to a value of only 0.038
which means that double and multi-component FIRST sources are shown
to have a little clustering amplitude beyond that angular limit,
i.e., intrinsic correlated double and multi components mostly
clustered under that scale. We also note that the physical size for
the $6^{\prime}$ angle at z=0.05 is $\sim 353$ kpc, approximately to
the scale of most radio jet and lobe ($\sim100$~kpc, Readhead et al.
1988; Jackson 1999, krolik 1999 ). All quasars in the SDSS DR3 have
redshift above 0.05. So we use the $6^{\prime}\times6^{\prime}$
FIRST map surrounding the quasar to determine possible complex
structure.

\begin{figure}[Ht]
\centering\includegraphics[height=6cm,width=8cm]{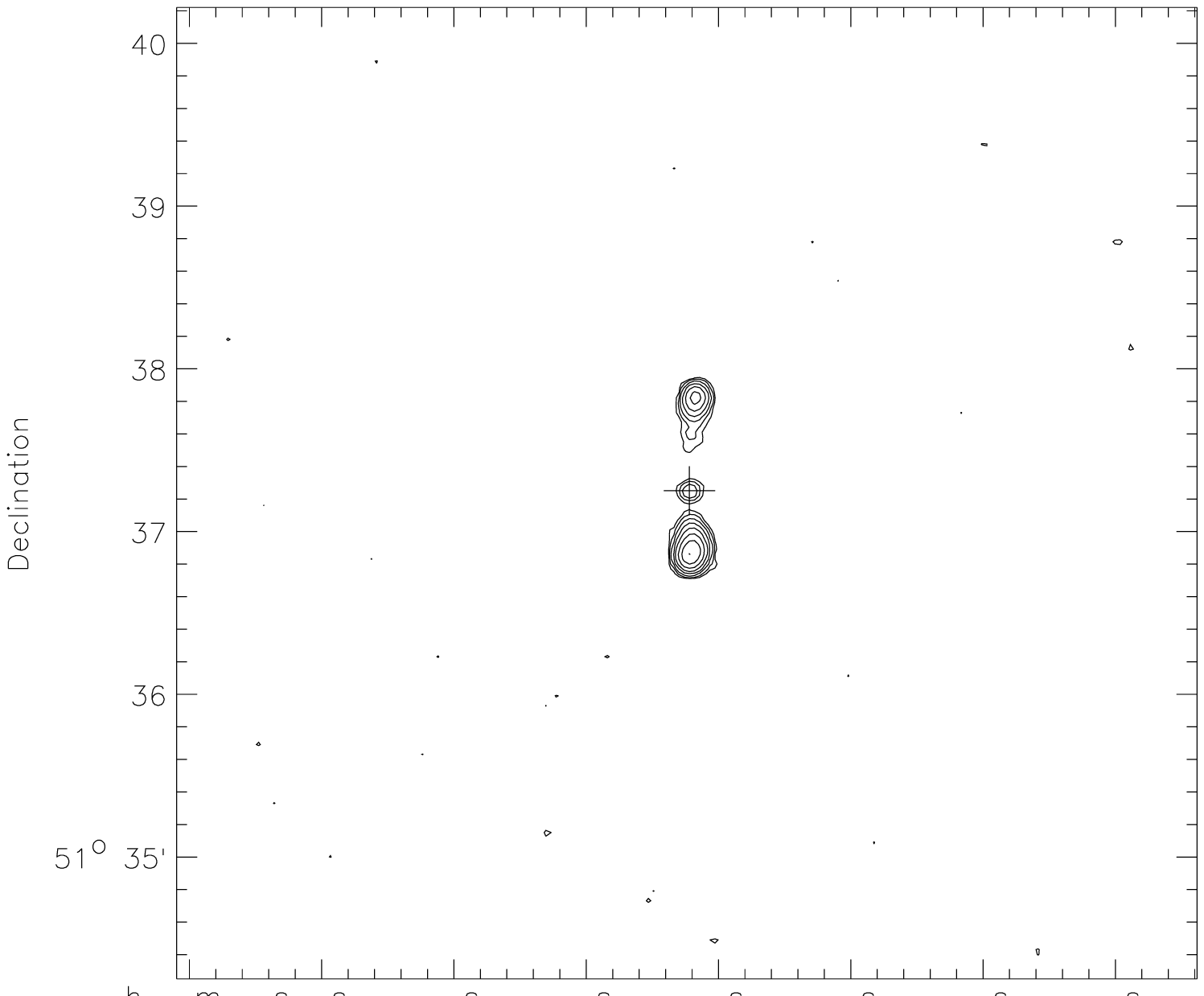}
\vspace{5mm}
\centering\includegraphics[height=6cm,width=8cm]{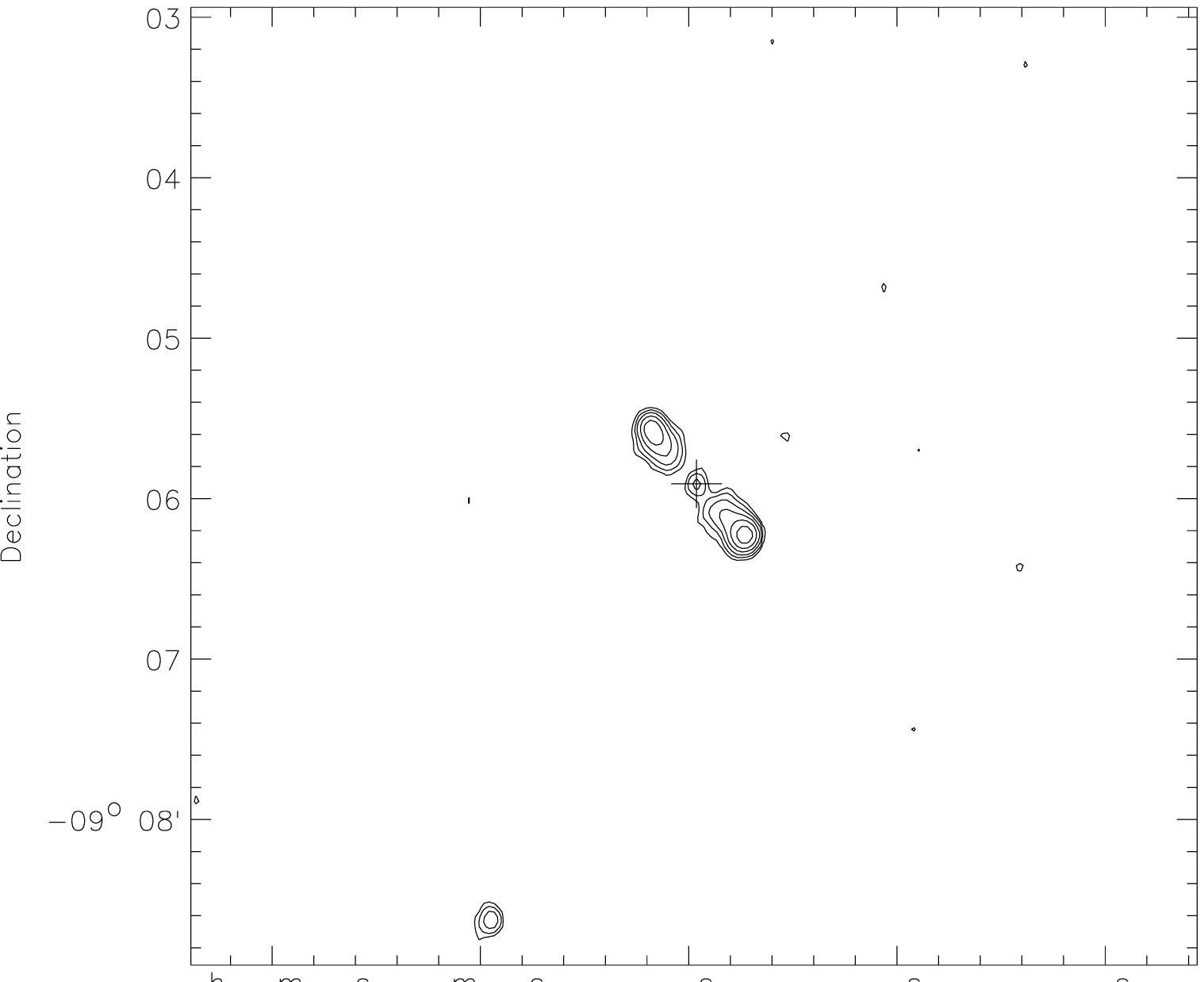}
\vspace{5mm}
\centering\includegraphics[height=6cm,width=8cm]{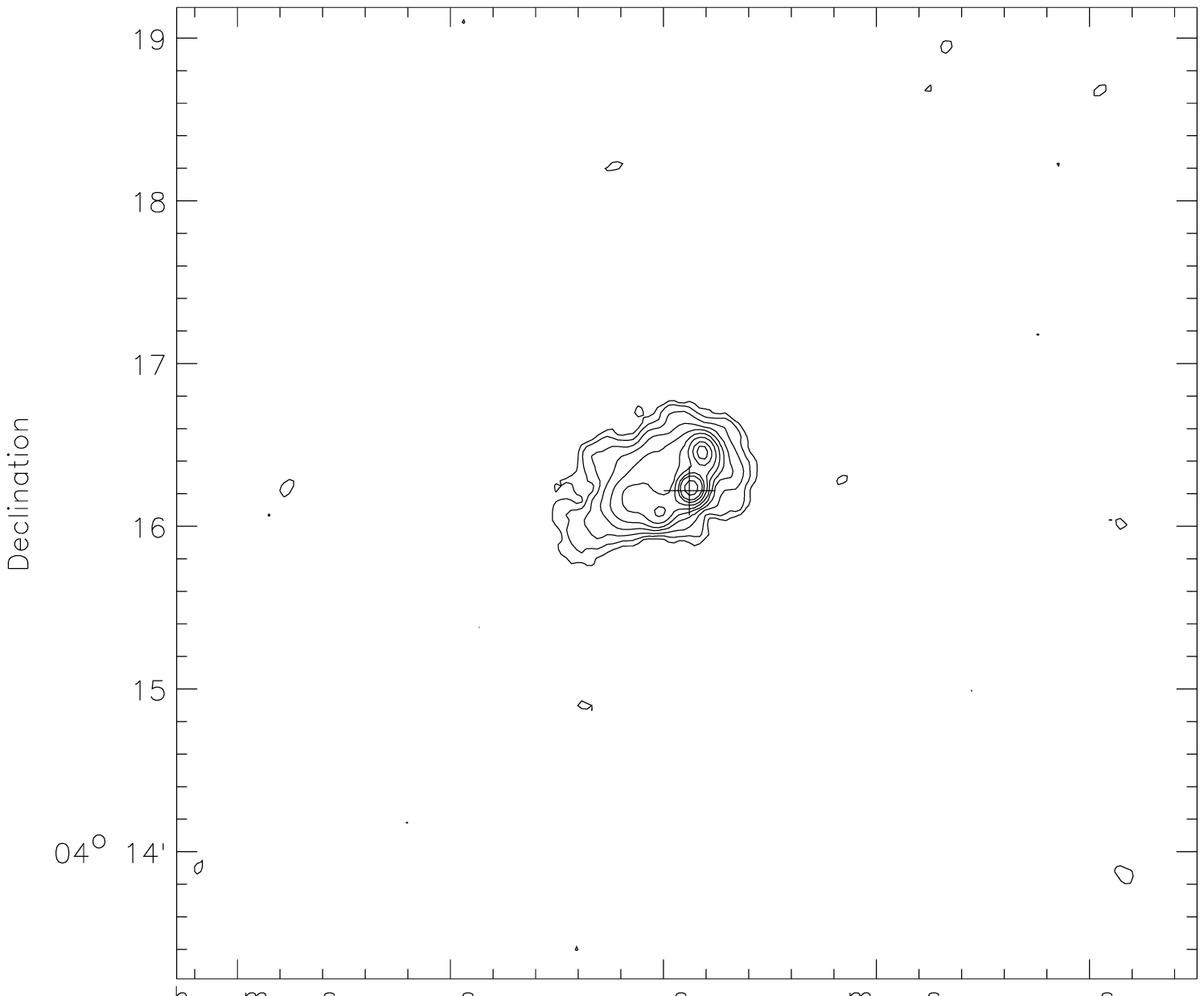}
\vspace{5mm}
\centering\includegraphics[height=6cm,width=8cm]{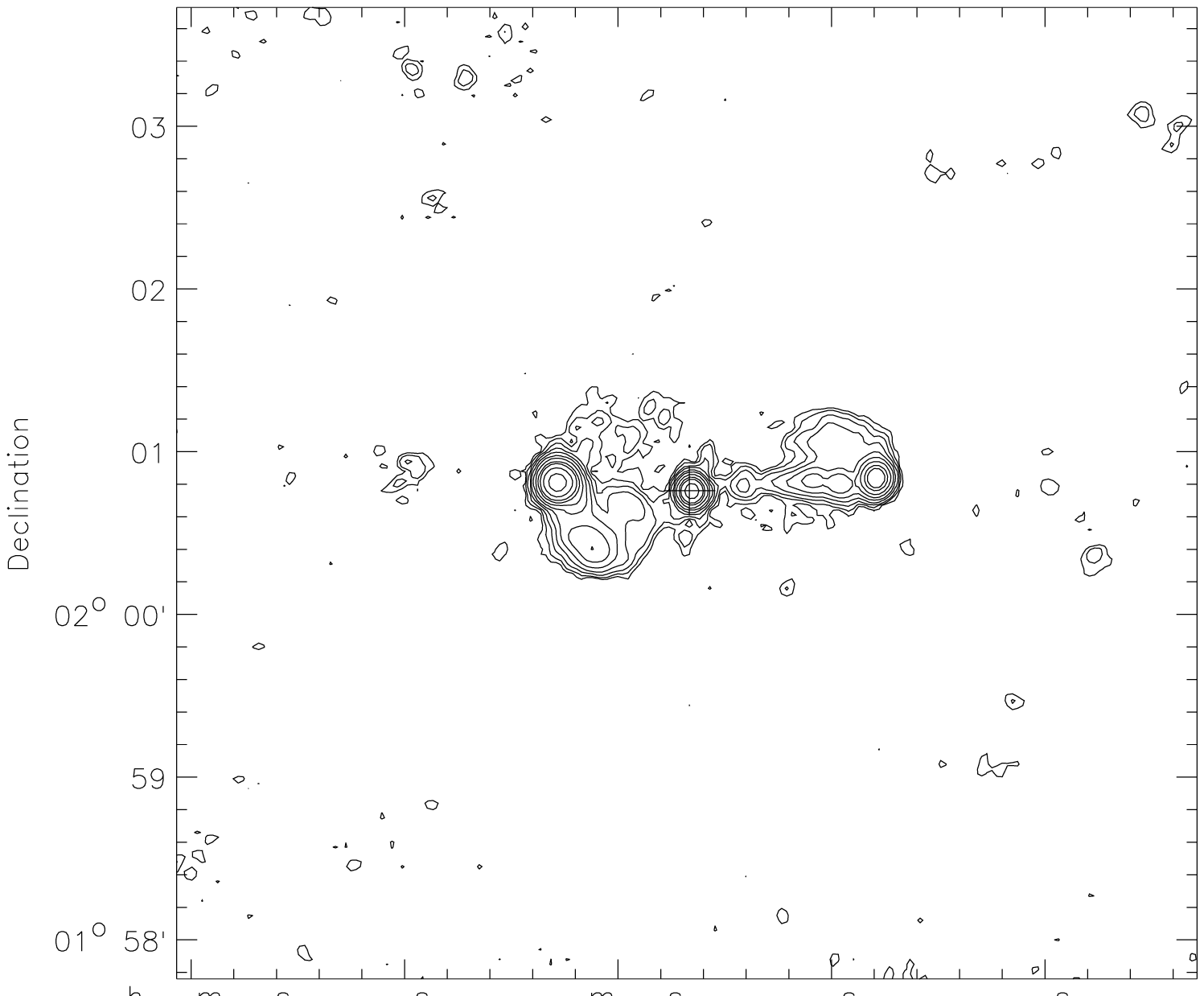}
\vspace{5mm}
\caption{\label{FIRST_image} Top two panels: FIRST
images of typical FRII-type radio quasars. Bottom two panels: the
FIRST images of the multi-component radio quasars.}
\end{figure}

The extended radio quasars are extracted in two steps. First, the
candidates of radio quasars are selected with one of the following
two simple criteria: (1) two radio sources locate nearly
symmetrically around the quasar position, i.e., the angle between
optical-radio connections lies $150^{\circ} \lesssim \theta \lesssim
210^{\circ}$, and the ratio of their distances to the quasar is $1/3
\lesssim d_1/d_2 \lesssim 3$; (2) more than two radio sources
scatter around the SDSS quasars. The first criterion allows us to
detect radio sources with symmetric lobes, either of FR-II or FR-I
type (Fanaroff \& Riley 1974), which is similar to that used by de
Vries et al. (2006). The FR-I sources are core-dominated sources
with a bright nucleus and two extended lobes with the surface
brightness decreasing towards the edges. In contrast, the FR-II
sources are generally lobe-dominated sources, and always show
brighter lobes, usually with a hot spot,  and with or without a weak
core (see Fig. \ref{FIRST_image}). However, when viewed at an
extreme angle, the FR-II sources may be dominated by the brighter
core (Barthel 1989; Hoekstra et al. 1997; Hardcastle et al.1998). By
the first criterion, sources with distorted asymmetric lobes may be
missed. The second criterion is designed for more complex radio
morphologies, such as sources with distorted asymmetric lobes and a
compact core or for cases in which extended lobes are resolved into
complex structure in the FIRST image. (see Fig. \ref{FIRST_image})
With these criteria, we selected 3115 radio quasar candidates. Among
them, 1035 sources are selected by the first criterion, and 2080 by
the second one. A $6^{\prime} \times6^{\prime}$ cutout of FIRST
image\footnote
{$http://archive.stsci.edu/vlafirst/getting\_started.html$} centered
at the quasar candidate was extracted for each SDSS source, and
visually checked. We use the radio morphologies in the 3CR radio
sources \footnote{$http:/www.jb.man.ac.uk/atlas/sample.html $} as
the reference for the true matches. We found about 70\% of these
radio components in $6^{\prime} \times6^{\prime}$ cutouts are likely
not related to the SDSS quasars, i.e., they are isolated components
or radio components related to other SDSS sources, or they have no
convincing evidence for their connection to the quasar.
 They will be excluded either from the radio flux estimation
or from the sample.

In the end, 859 extended sources are selected (with unambiguous
radio lobes, jets) from initial 3115 candidates. Among them, about
half (409) show FR-II morphologies, and 564 have radio core
components (within 2\texttt{"} circle of optical quasar position) in
the FIRST catalog. In comparison, de Vries et al. found 422 FR-II
quasars using Abazajian et al. (2005) DR3 quasar sample (44984
sources),
 of which 359 are in common. The members of radio counterparts in
the de Vries sample that are not included in our sample are 63. Most
these ``lost'' objects (41 out of 63) are not included in the SDSS
DR3 quasar catalog of Schneider et al. (2005). The other 18 were
excluded by us due to their "FIRST-only" or "ROSAT-only"
target-flag. The remaining 4 were excluded by us for their
inconvincible connection between the radio and optical sources. We
also notice that, for 83 of the 359 common quasars in both sample
that show more complex radio structure than simply "double-lobes",
i.e., the lobes resolved in the FIRST image, we account for the
radio fluxes of all components. As a comparison, Vries et al. (2006)
only considered the radio fluxes of the two components located
symmetrically around the quasar, thus the total radio fluxes are
systematically underestimated for these sources. The flux difference
between the two ranges from $2\%$ to $86\%$ , with a typical value
of $31.2\%$.

The maximum projected physical distance of the lobe component is
independent of the radio power (see left panel of Fig.
\ref{lobe-distrib}) while the rest-frame peak intensity increases
with radio power as shown in right panel of Fig. \ref{lobe-distrib}.
The former is expected if the radio power does not vary significant
during the quasar activity phase while the lobe to push further
outward. The latter may result from the interaction of powerful
radio jet with medium, which will produce more relativistic
electrons, thus high intrinsic brightness.

\begin{figure}
\plottwo{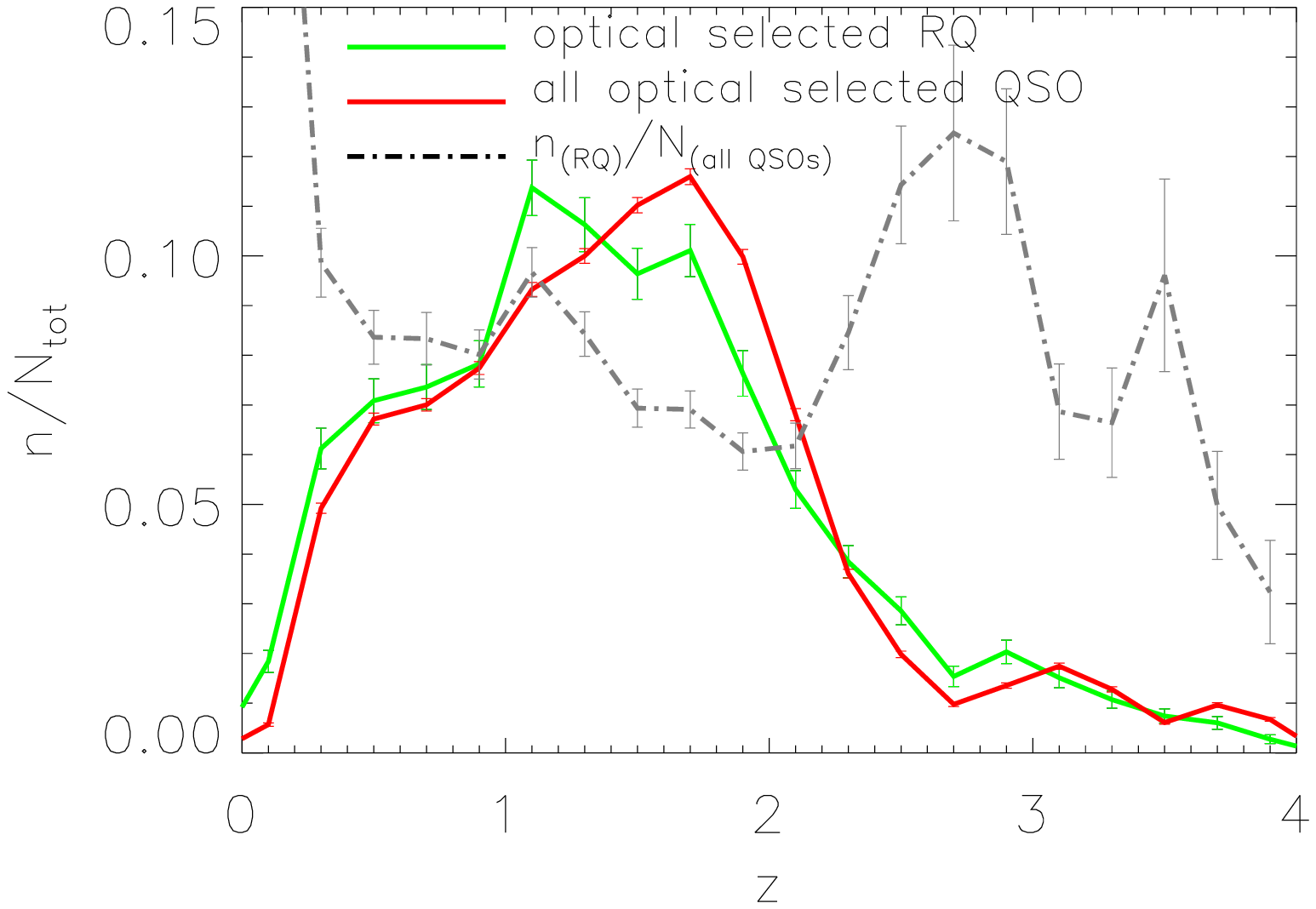}{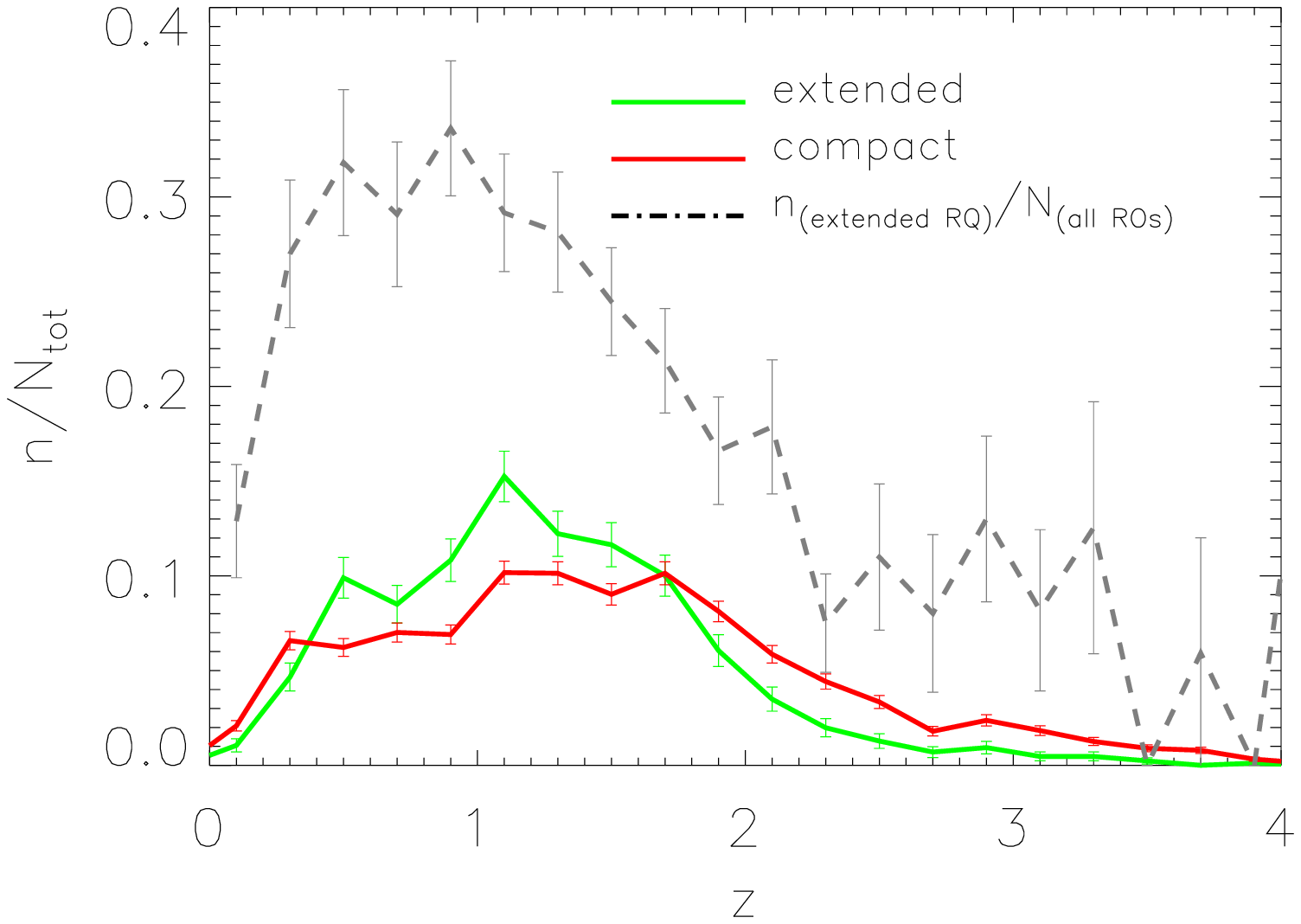}
\caption{\label{redshift} Left
panel: The redshift distribution of optical selected radio quasars
in our sample (green line) and of all optical selected DR3 quasars
in Schneider et al. (2005) (red line). The dot-dashed line shown the
ratio of optical+radio quasar to all optical selected quasars. Right
panel: The redshift distribution of extended radio quasars (green
line) and compact quasars (red line). And the dot-dashed line shown
the fraction of extended quasars to all radio quasars. Obviously the
ratio drops as redshift increases up to z=2.5. }
\end{figure}

By merging the compact and extended radio quasar sample after
eliminating duplicate sources, we obtain a sample of 3641 radio
quasars. Among them, 859 are multi-component sources (extended
sources, hereafter) and the remain 2782 are single component sources
(hereafter compact sources). The average apparent magnitude of this
sample is $m_{i} \sim 18.78$, the median redshift is $\sim 1.36$
(see Fig.\ref{redshift}). The distributions of optical selected and
optical+radio selected sub-sample are significant difference. As
noticed by Richards et al. (2006), radio-selected and
optically-selected quasars show different redshift distribution in
the sense that the optical selected quasars show deeper deficit at
$z\sim 2.7$, while the optical+radio quasars distribute smoother at
$z>2$. This is due to the selection effect of the optically-selected
SDSS quasars based on optical colors. We can also see that radio
quasars peak at a slightly lower redshift and are more abundant
between redshifts of 2.2 and 3.0. This may be due to the
comparatively shallower survey depth of FIRST, so that we can find
more radio counterparts of optical-selected quasars at lower
redshift.

Note that the redshift distribution of extended sources peaks at
lower redshift than the compact quasars, i.e., the fraction of
extended sources decreases with redshift. The median radio flux is
5.49 mJy for compact quasars and 48.31 mJy for extended quasars.

\section{On the selection effect of radio quasar sample}

\subsection{The incompleteness of extended sources caused by the surface
            brightness limit}

FIRST is not sensitive to low surface brightness emission due to its small
beam size. As such diffuse emission, if present, will be missed in the FIRST
survey. It should be noted that the surface brightness are lowered due to
cosmological expansion ($I\propto(1+z)^{-4-\alpha_r}$), as such lobes may
fall below the detection limit (0.75 mJy per beam) of the FIRST survey at
large redshift (see below). In the worst case, if the source is dominated
by a diffuse component, they may escape the detection at all in the FIRST
survey. We use cross correlation between SDSS and NVSS to constrain this.

The 1.4GHz NRAO VLA Sky Survey (NVSS; Condon etal 1998 ) provided a
catalog contain $2\times 10^6 $ sources stronger than $2.5$ mJy. It
is $90$ percent complete at integrated flux density $S_{1.4GHz}=3$
mJy and $99$ percent complete at $S_{1.4GHz}=3.5$ mJy. With a
synthesized beam of $45^{''}$ (FWHM), NVSS is much more sensitive to
lower surface brightness components than FIRST. Therefore, it can be
used to check the fraction of quasars with only diffuse emission.
For typical redshift of quasars in this sample, the radio emission
should be unresolved by NVSS.

In the following we will consider radio sources with NVSS flux
density larger than $3$ mJy. The matching radius used in the
cross-correlation of DR3 quasars and NVSS sources is trade-off
between random contamination and completeness. First, we estimate
that the fractions of chance coincidence corresponding to $15^{''}$,
$20^{''}$, $25^{''}$ matching radius are $\sim 3.95$\%,
$\sim6.68\%$, $\sim9.64\%$ respectively. Second, we checked 775
extended sources with FIRST flux above 3~mJy in our FIRST-DR3 quasar
sample. Among them, NVSS counterparts are found for 719 quasars ($
\sim 92.8$\%) within $20^{''}$ offset of the quasars. The fraction
decreases to $87.65\%$ with a $15^{''}$ matching radius, and
increase to $95.6\%$ with $25^{''}$ matching radius. We take
$20^{''}$ as the matching radius for the NVSS-SDSS match.

\begin{figure}
\plottwo{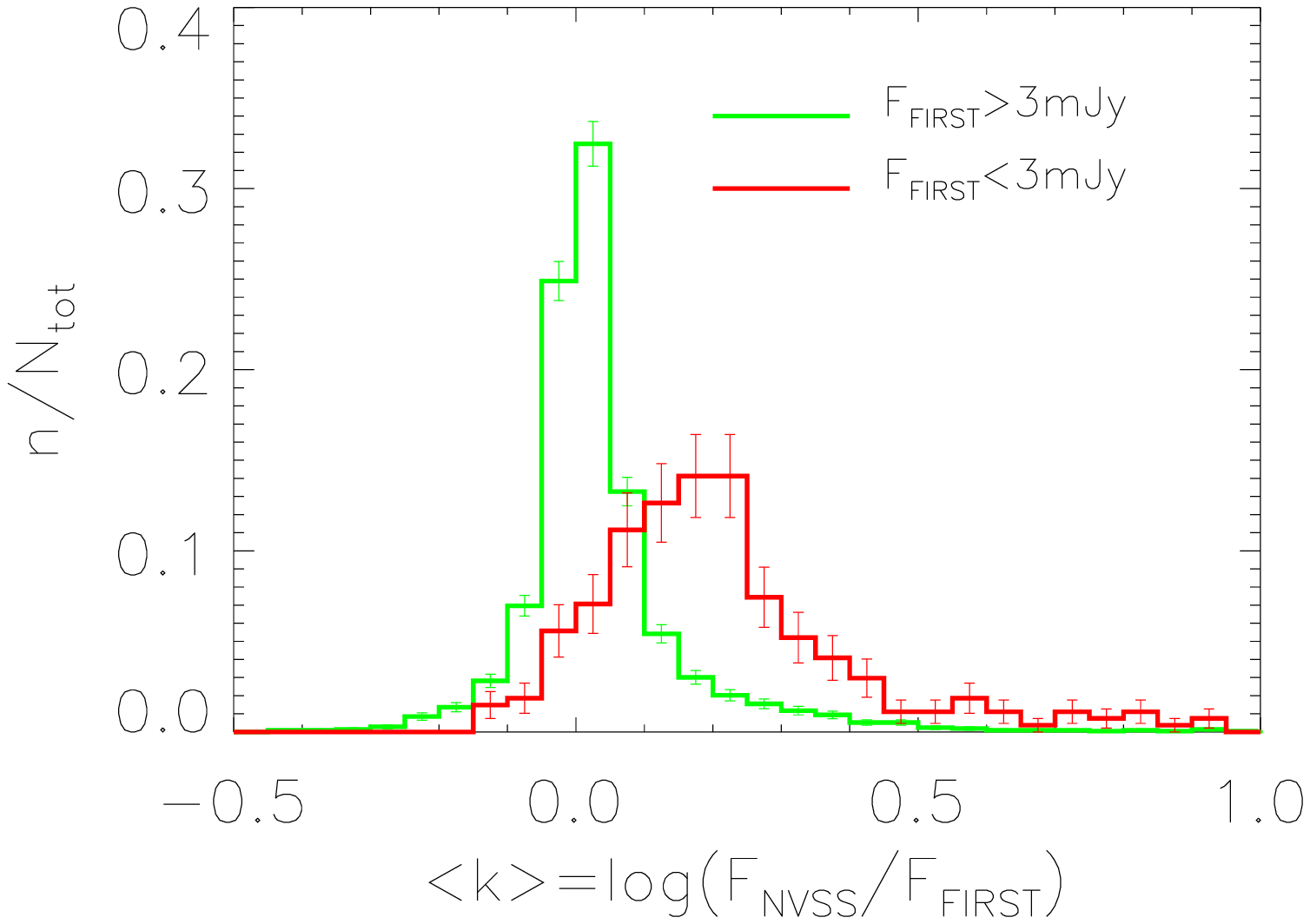}{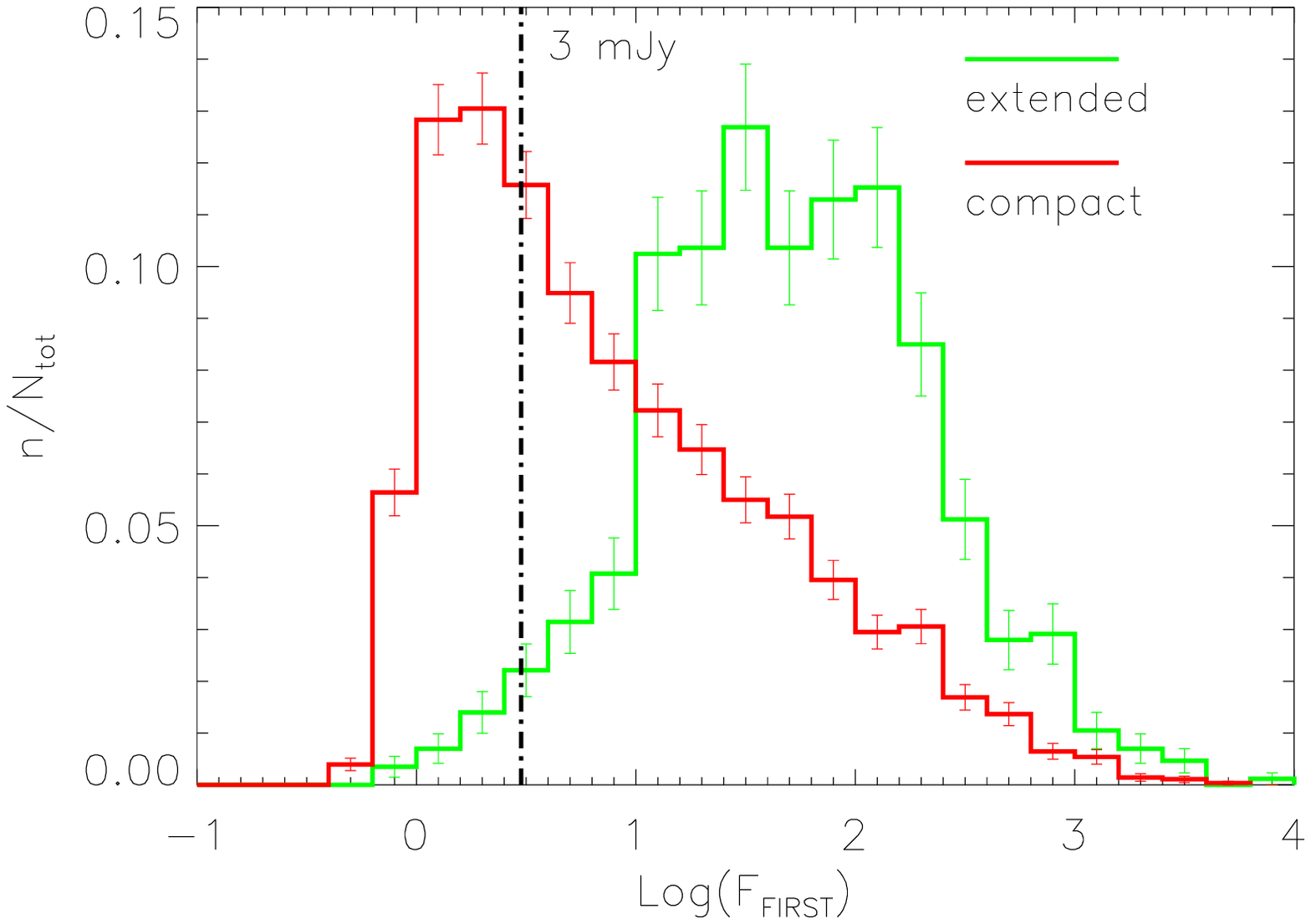}
\centering\includegraphics[height=6cm,width=8cm]{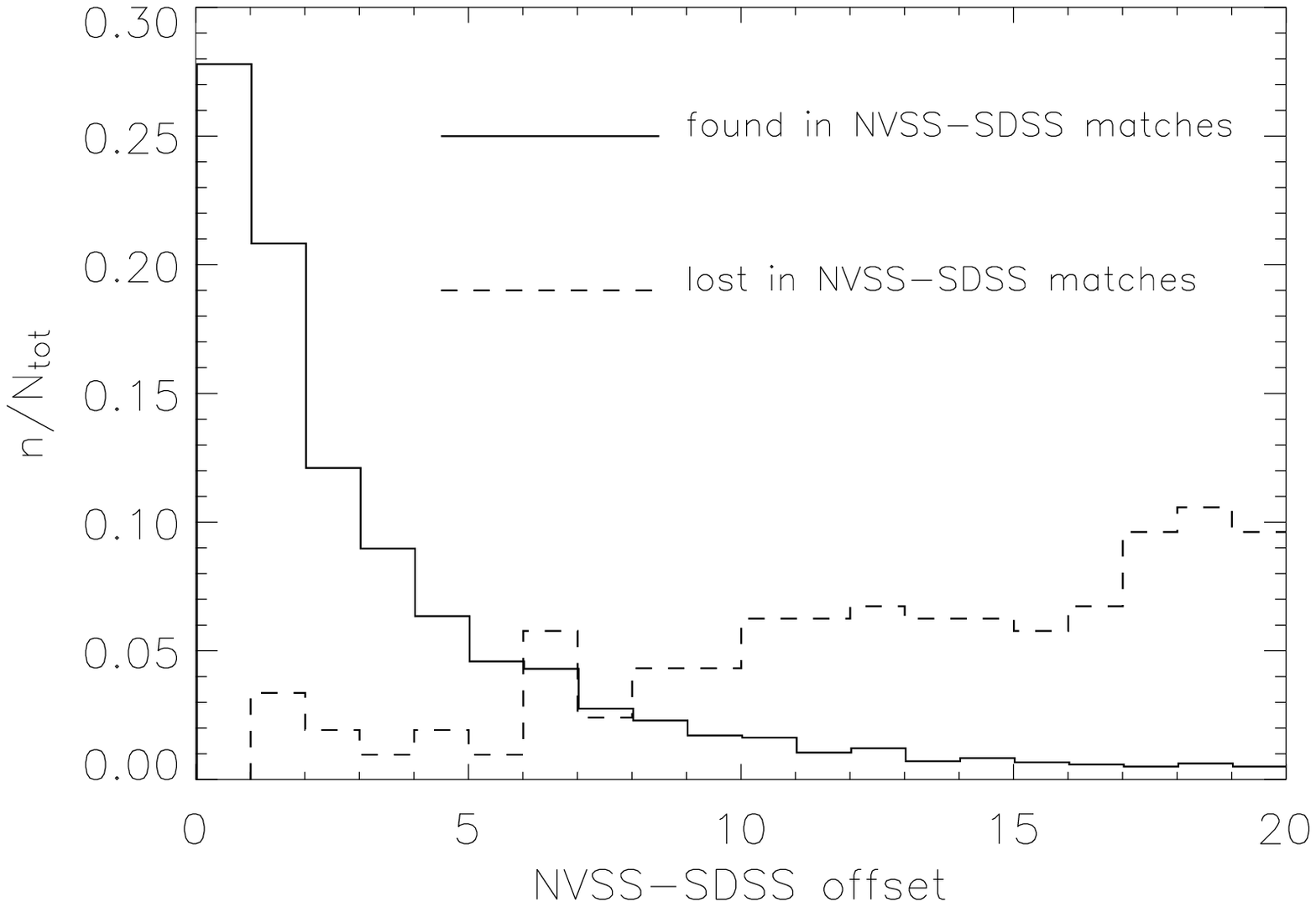}
\linespread {0.3} \caption{\label{NVSS-FIRST_flux} Top left panel:
The distribution of log$(f_{NVSS}/f_{FIRST})$ . The green line
represents the sources with FIRST flux above the 3 mJy limit, and
the red line with FIRST flux below the 3 mJy limit. The FIRST flux
is the sum of all radio sources within NVSS beam size. Top right
panel: The flux distribution of extended sources (green line) and
compact sources (red line). The vertical dot-dash line marks the 3
mJy flux. Bottom panel:  The NVSS-SDSS offset distribution of the
"selected" matches and "rejected" matches in 20\texttt{"} matching
radius.}
\end{figure}

With $20^{''}$ matching radius, we extracted $3029$ NVSS radio
counterparts to SDSS DR3 quasars in the overlapping area of FIRST,
SDSS DR3 and NVSS, with the NVSS flux density limit of 3 mJy. In
this DR3-NVSS sample, $227$ sources are not present in our DR3-FIRST
radio quasar sample. $208$ of these "lost" sources
 locate at a distance larger than the NVSS major axis from the quasar.
We found the NVSS-SDSS offset distribution of these "lost" sources
 are significantly different from the "found" ones
(see the bottom panel of Fig. \ref{NVSS-FIRST_flux}),
 indicating most of them are due to chance coincidence.
This number is close to the expected rate of chance coincidences
($\sim 6.68\%$). Note in passing four sources are found in our
DR3-FIRST sample, but their NVSS flux density are less than 3 mJy.
Therefore, no more than 19 extended radio quasars may be missed due
to the higher FIRST resolution at the NVSS flux density limit of 3
mJy. This lost fraction is $\sim 19/(775+19)\sim2.3\%$. Therefore,
above 3 mJy flux density limit, the radio quasars with only diffuse
radio sources are rare.

However, weak diffuse emission is likely to be overlooked by FIRST.
We plotted the distribution of $k=\log(f_{NVSS}/f_{FIRST})$ at FIRST
flux $>$ 3~mJy and $<$ 3~mJy (left panel of
Fig.\ref{NVSS-FIRST_flux}), where $f_{FIRST}$ is the sum of all
radio sources within the NVSS beam size. The $k$ is peaked at zero
with a tail toward $<k>\sim 0.04$ above 3 mJy, which suggests that
most of the NVSS flux had been detected by the FIRST. But below 3
mJy limit, $<k>$ increased to 0.24. This indicates that some diffuse
radio flux such as weak lobes of extended sources can be
underestimated by the FIRST as flux density decreases. As a result,
some extended sources with low surface brightness may be
mis-identified as compact sources. This may explain why most of the
extended sources have flux greater than 3 ~mJy (see the right panel
of Fig.\ref{NVSS-FIRST_flux}).

\subsection{The selection effects of a radio quasar sample}

In this section, we will address the selection effects introduced by
the positional coincidence alone using our radio quasar sample. This
includes the lost fraction of lobe-only objects and the
underestimate of the extended flux. In the following, we will divide
the radio quasars into extended and compact sources according to
whether one or more extra-core components are present or not. The
core component is defined as the single component within the
$2^{''}$ radius of the optical quasar.

\begin{figure}
\plottwo{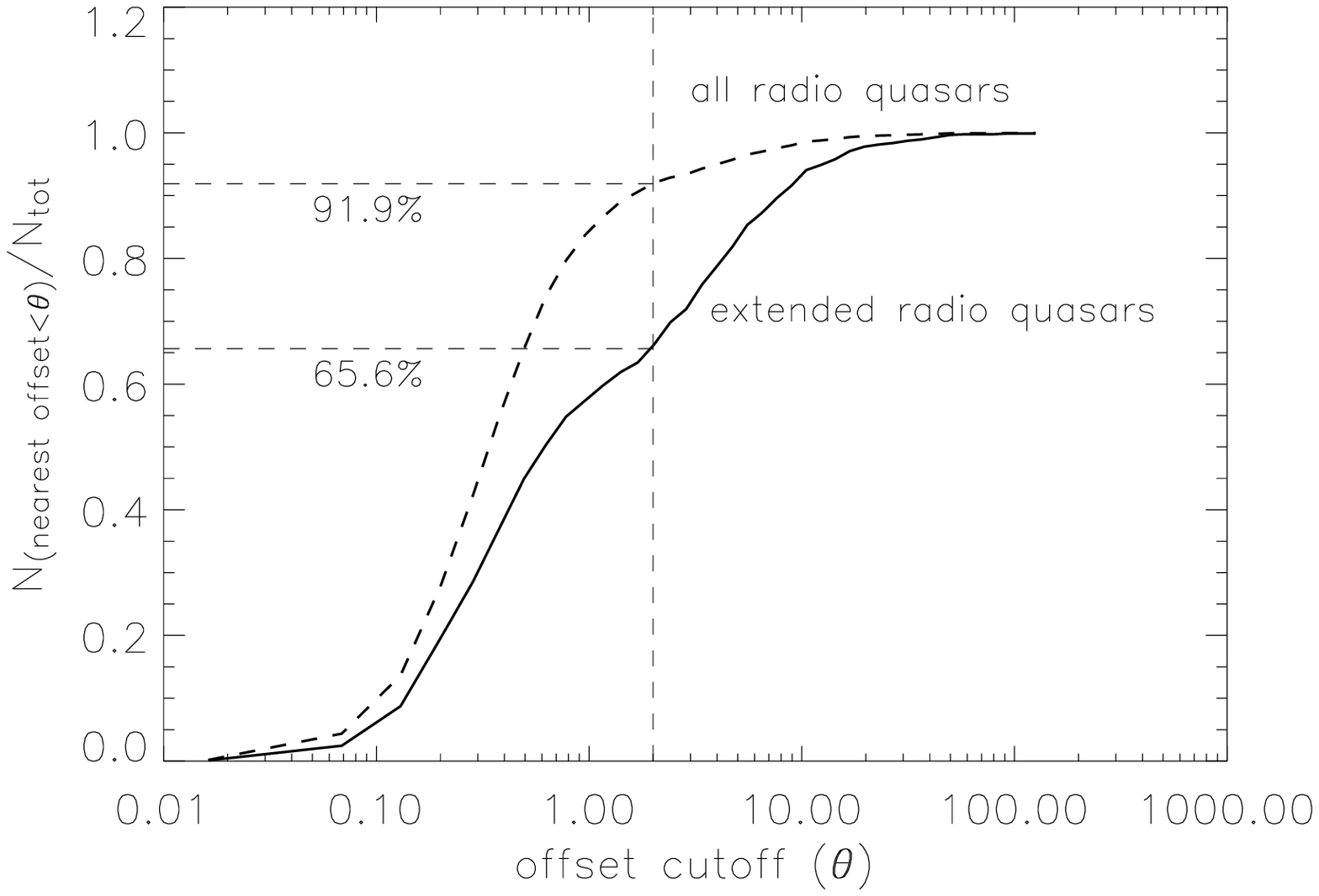}{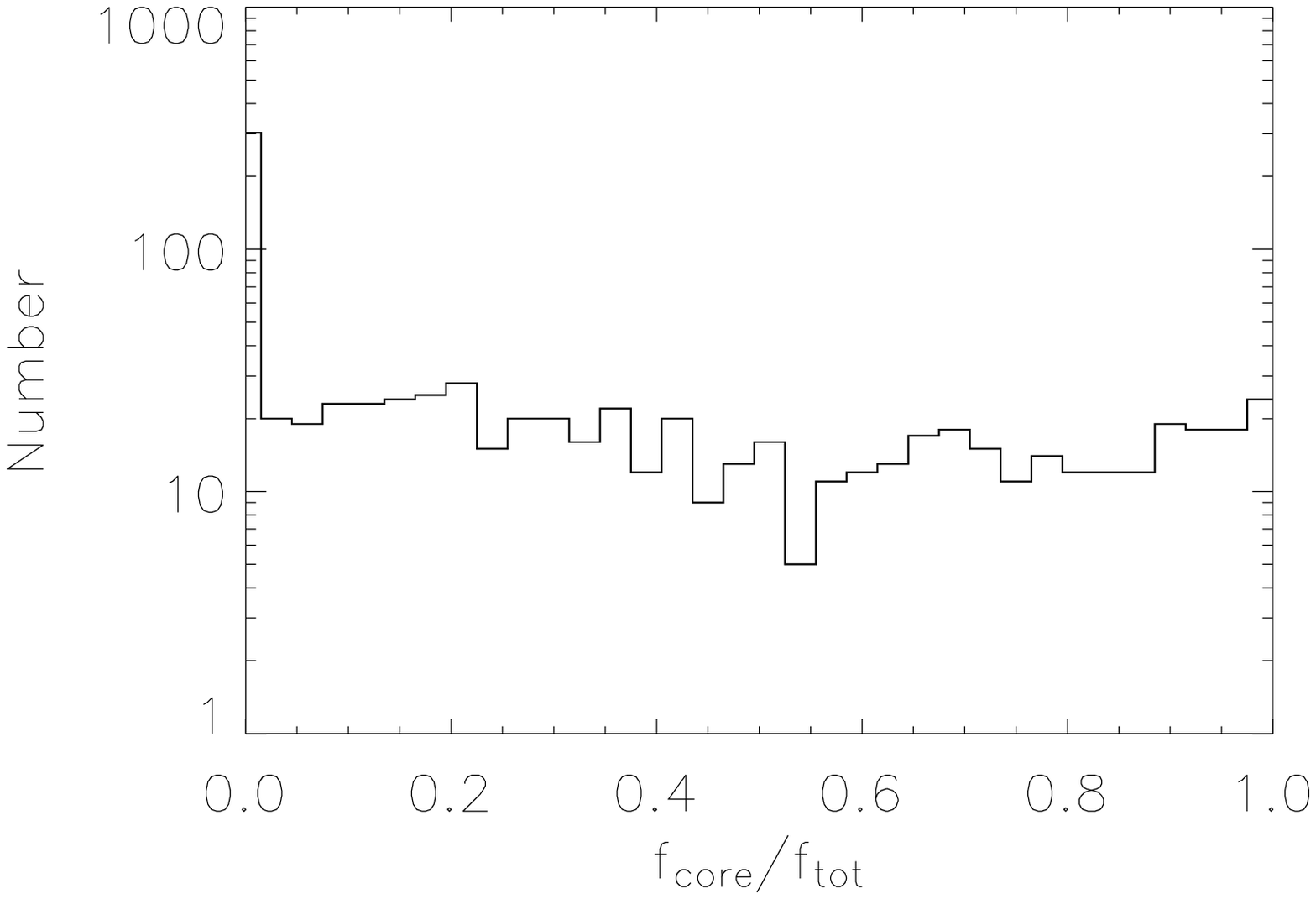} \caption{\label{obs-e-c-chars}
Left panel: The accumulated fraction of sources with nearest FIRST
counterpart within certain matching radius. The solid line
represents the lobe-dominant sources and the dashed one all the
radio quasars. Right panel: The distribution of flux ratio $q$. $q$
defined as the flux ratio of "core" component (the radio counterpart
within 2\texttt{"}) to the total radio flux (summation over all
radio counterparts that associate to the radio quasar). }
\end{figure}

The distribution of position offsets of the closest radio component
to the SDSS quasar is shown in the left panel of
Fig.\ref{obs-e-c-chars} for our radio quasar sample. We find that
13.0\% of quasars will be lost with a matching radius in the
position offset of $1\farcs2$, 10.4\% at $1\farcs5$ and $8.1\%$ at
$2^{''}$. With these numbers, we conclude that the fraction of radio
quasars missed due to lack of detectable radio cores is low.

Using the positional coincidence will underestimate the radio flux
density if there is one or more off-core components even if a radio
core is present. This will be important particularly in
lobe-dominated quasars. To quantify this bias, we calculate $q$ as
the flux ratio of the core component to the total radio flux (the
summation over all radio counterparts that are associated with the
radio quasar). The distribution of $q$ for 564 quasars with
core-lobe structure and 295 objects without cores ($\sim 35\%$ of
all extended sources) are shown in Fig.\ref{obs-e-c-chars}. It is
interesting to note that the distribution of $q$ keep nearly
constant between $0 < q < 1.0$, i.e., the number of strong-lobe
sources ($q < 0.5$) and weak-lobe sources ($q > 0.5$) are similar.
We found that the radio fluxes are either underestimated by a factor
more than two or complete missed in about 16.8\% of all radio
quasars. The sources with two or more components are significantly
more numerous than estimated by Ivezic et al. (2002).

\begin{figure}
\plottwo{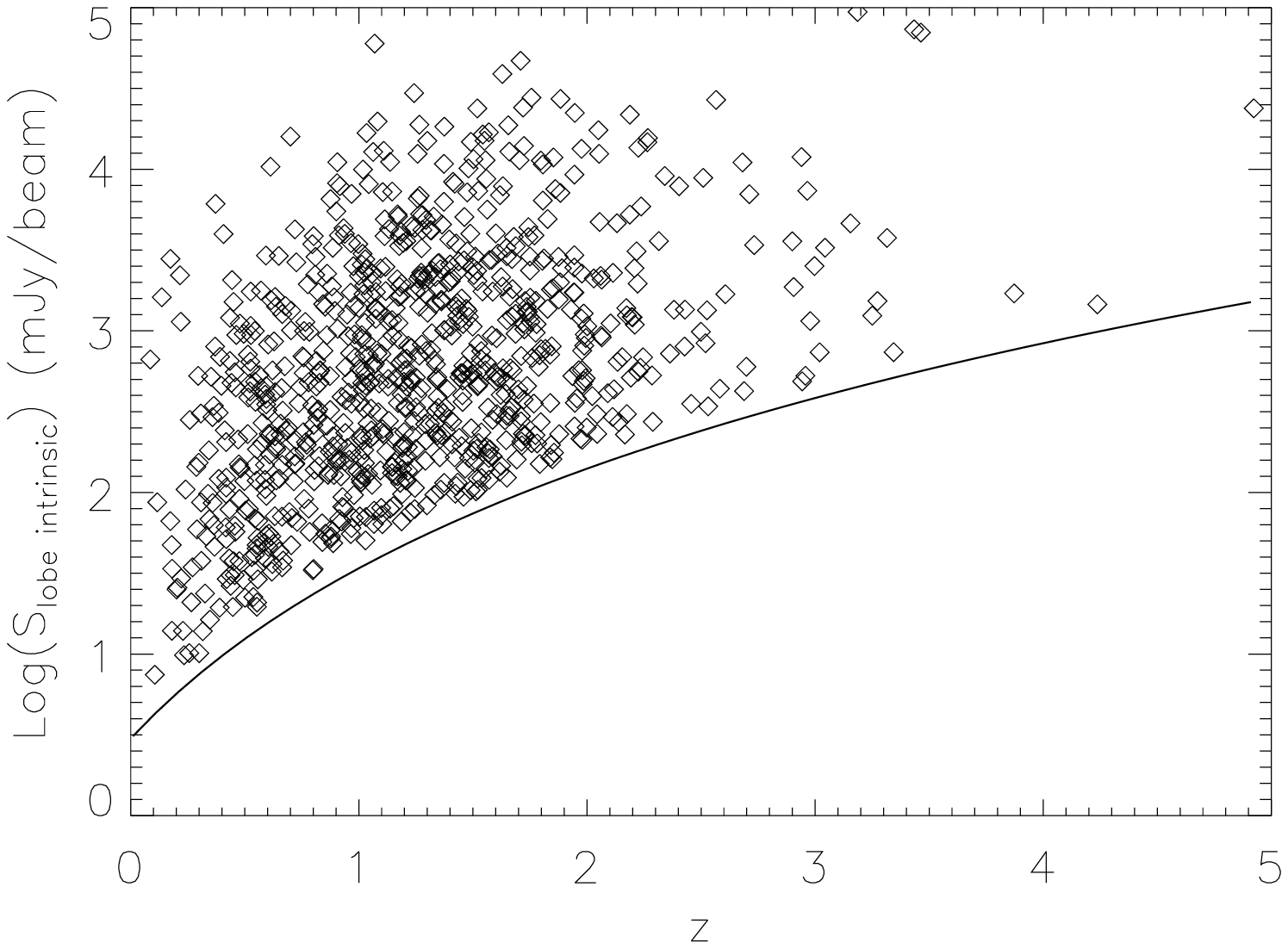}{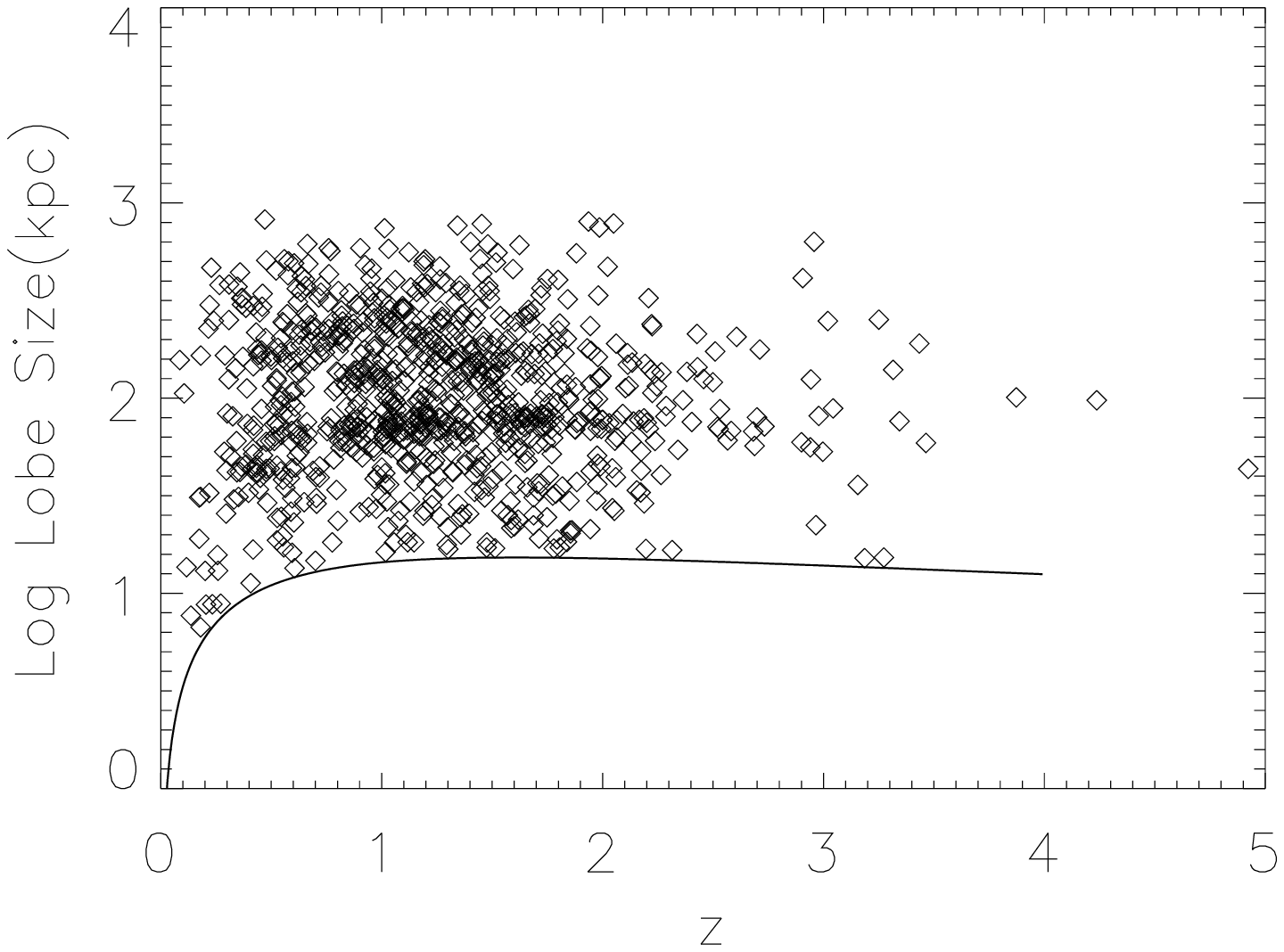} \caption{\label{lobe-z} Left
panel: The intrinsic brightness of the lobes versus redshift. The
solid line represents the surface brightness at the detection limit
as 0.75 mJy/beam. Right panel: The physical size of extended sources
at different redshift. The size is the physical distance of the
furthest radio component associated with the quasar. The solid line
marks the minimum distance that can
 be resolved in the FIRST survey at the corresponding redshift (on scales down
to $\sim 1/3$ the beam size of 5\farcs4.)}
\end{figure}

Note that with our definitions whether a quasar is compact or
extended depends also on the redshift of the quasar.
Fig.\ref{redshift} shows that the fraction of extended sources peaks
at redshift $\sim 1.0$ and decreases dramatically at redshift
$\gtrsim 2.0$. Note that this is not due to the increase of the
angular distance because it is peaked around 1.5 for the
cosmological model adopt here, and most extended sources at redshift
less than 0.5 have sizes that should be resolved at redshift around
2-4 ( Fig.\ref{lobe-z}). This can be due to either an evolution in
the radio structure of quasars or the decrease of brightness caused
by cosmology expansion ($I\propto(1+z)^{-4-\alpha_r}$)
(Fig.\ref{lobe-z}).

\begin{figure}
\centering\includegraphics[height=6cm,width=8cm]{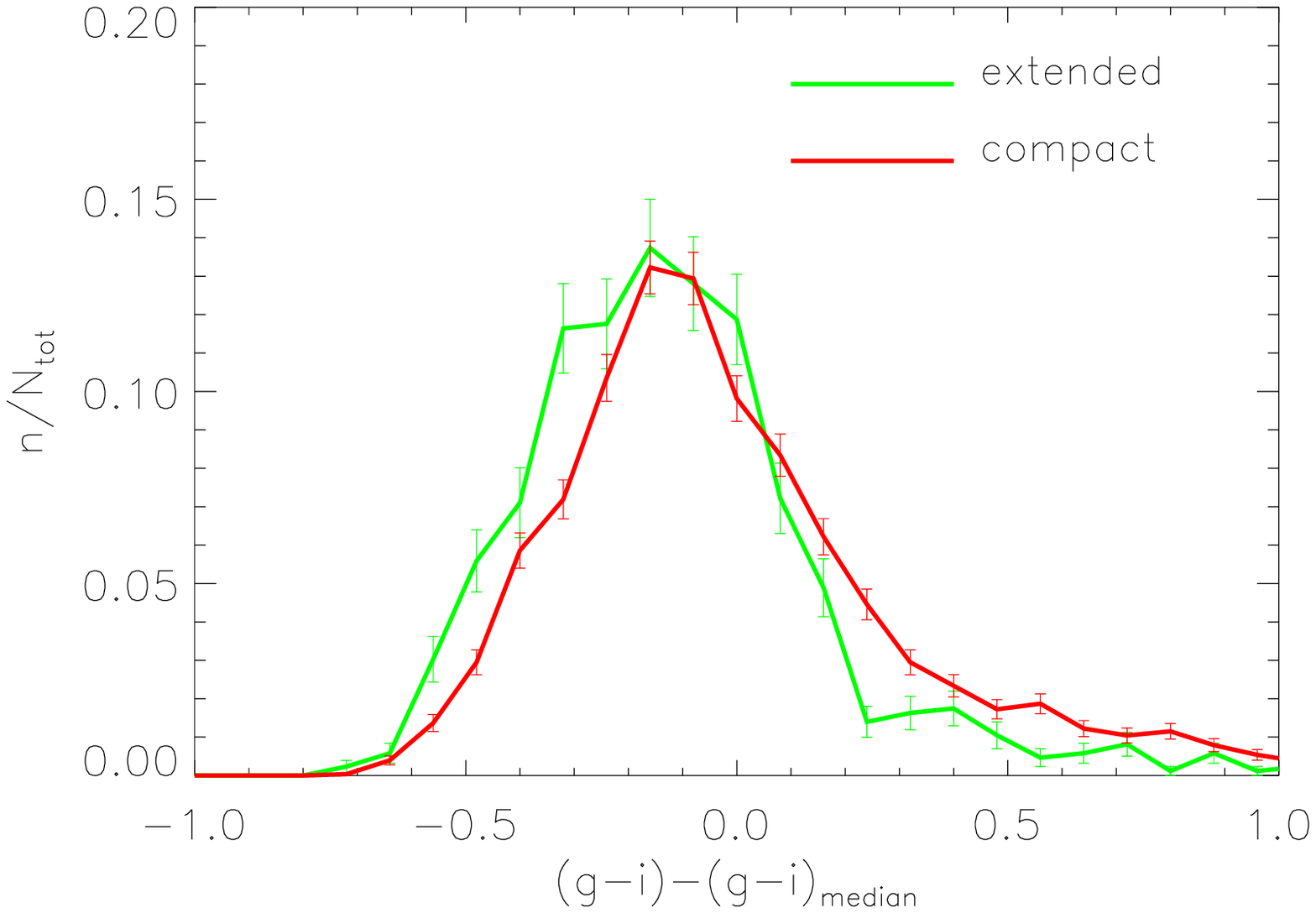}
\linespread {0.3}
\centering\includegraphics[height=6cm,width=8cm]{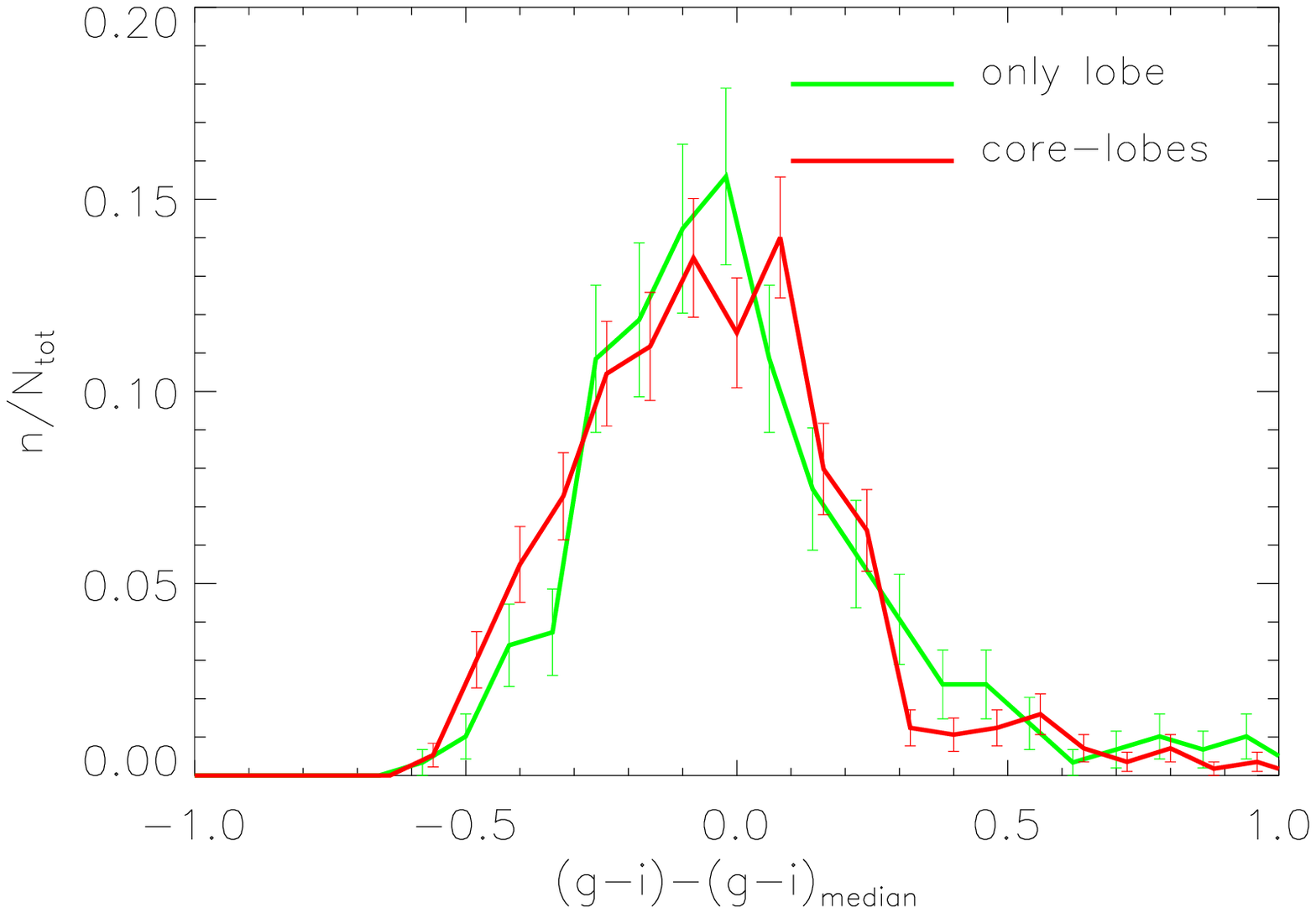}
\linespread {0.3}
\centering\includegraphics[height=6cm,width=8cm]{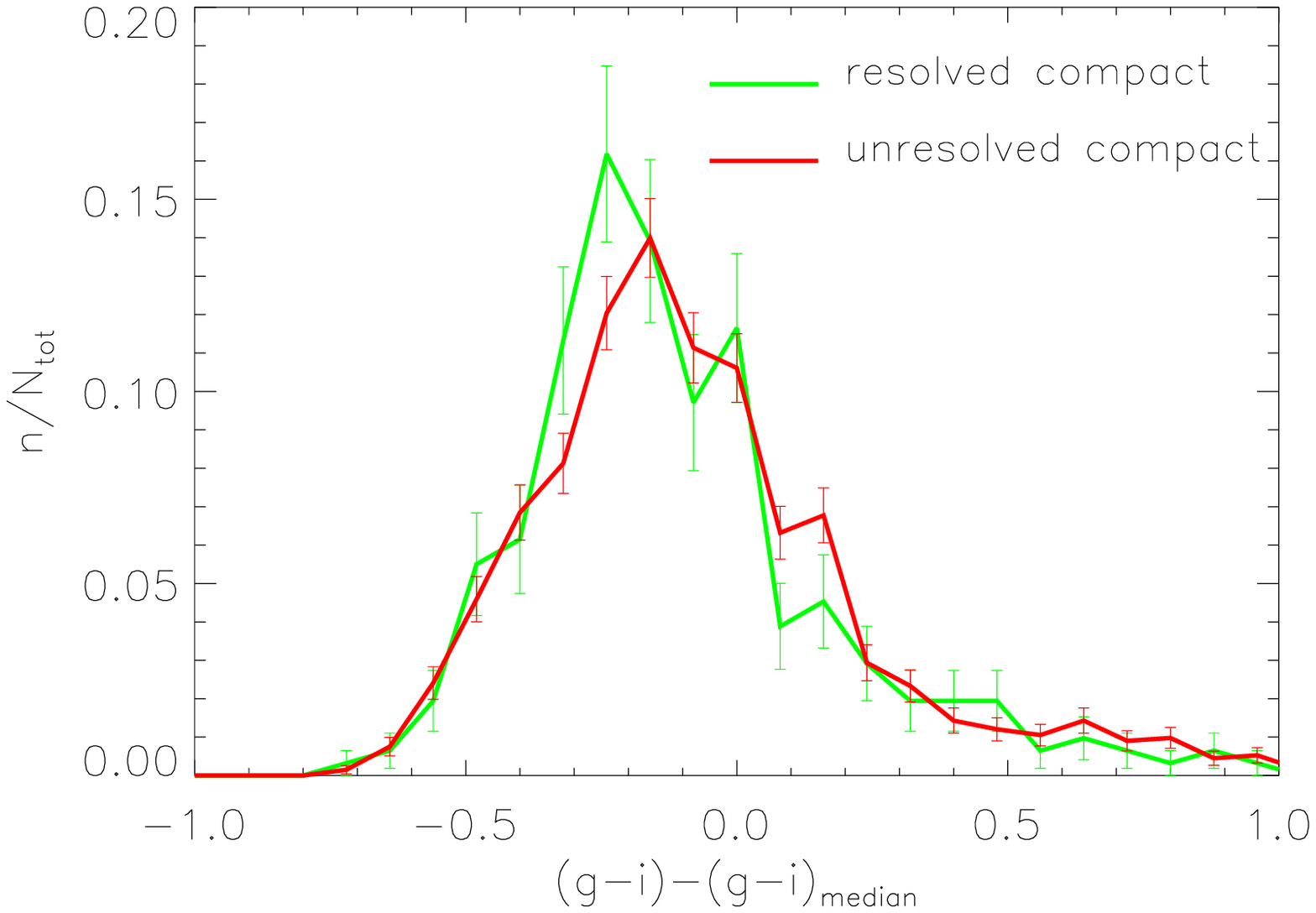}
\linespread {0.3}
\centering\includegraphics[height=6cm,width=8cm]{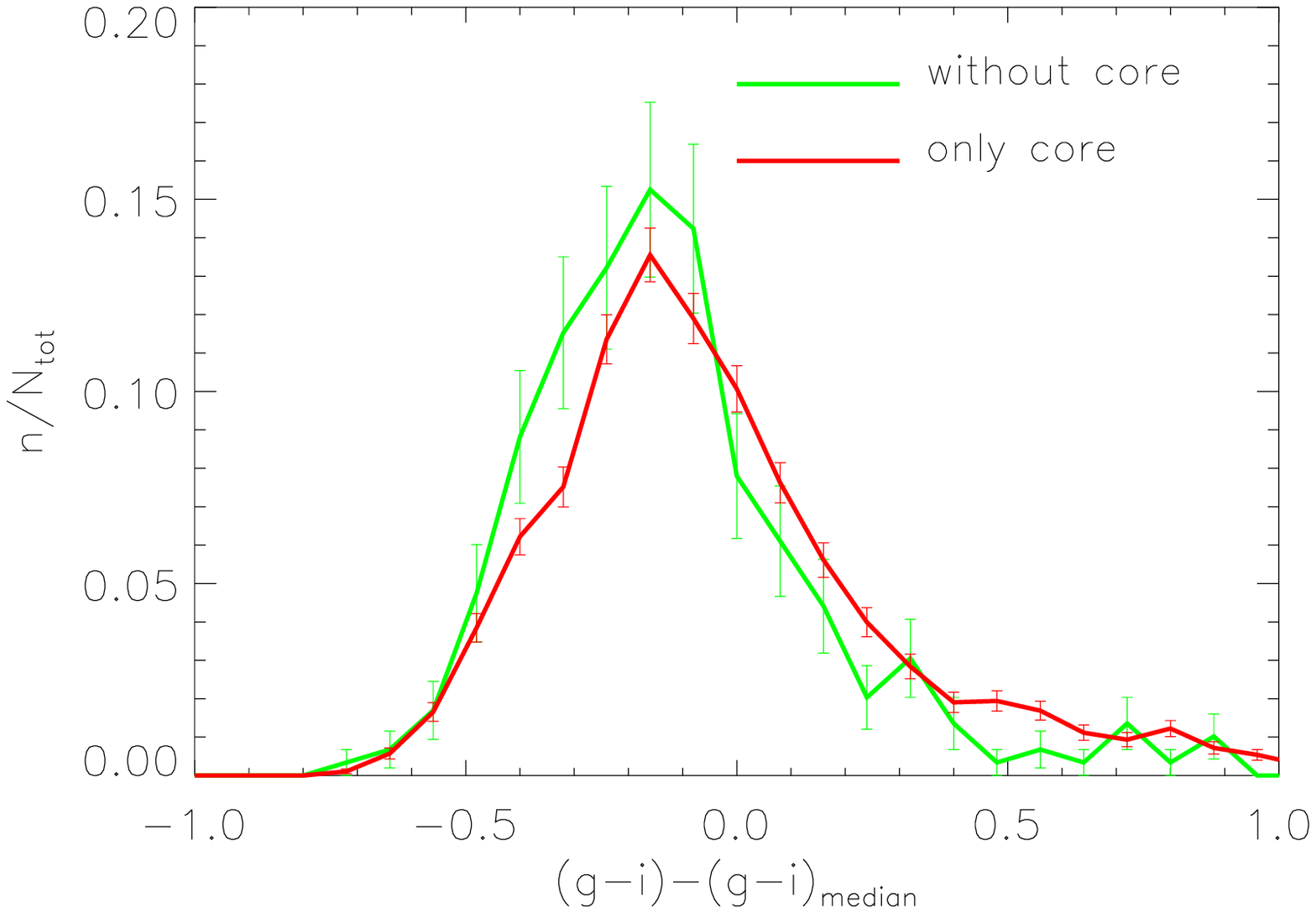}
\linespread {0.3} \caption{\label{color} The median-subtracted $g-i$
color distribution for quasars with redshift $0<z<3$ (refer to the
text for definition). A comparison between extended (green lines)
and compact radio quasars (red line) is shown in the top left panel;
between resolved (green line) and unresolved compact radio quasars
(red line) in the bottom left panel; between extended quasars with
(red line) and without (green line) core in the top right panel; and
between only-lobe (green line) and only-core (red line) quasars. }

\end{figure}

We show the optical color distributions in Fig. \ref{color}. In
order to eliminate the dependence on redshift, we divide our sample
into 20 redshift bins. For each bin, we subtract from $g-i$ the
median value of $(g-i)_{median}$ in that bin. Surprisingly, the
color distribution of extended radio quasars (with and without core)
does not appear redder, but slightly bluer than the compact radio
quasars. Kolmogorov-Smirnoff test deems the significant difference
between the color distribution of extended and compact sources at
the confidence level $\gtrsim 97\%$. Also, student-t test suggests
that the mean color of the extended radio quasars is bluer than that
of the compact quasars at a confidence level of $\gtrsim 99\%$.

The result remains the same for the core-only and lobe-only sources
and for quasars with unresolved and resolved cores (resolved core:
log$ (f_{int}/f_{peak})^2>0.1$, with $f_{int}>3.6$ mJy in order to
gain sufficient S/N ratio, (Ivezic et al. 2002). (see
Fig.\ref{color} ). But, the core-lobe sources and the only-lobe
sources are indistinguishable. These results are in line with that
of de Vries et al.(2006), who found that the composite spectrum of
FR-II quasars is flatter than that of radio compact quasars.

\begin{figure}[Ht]
\plottwo{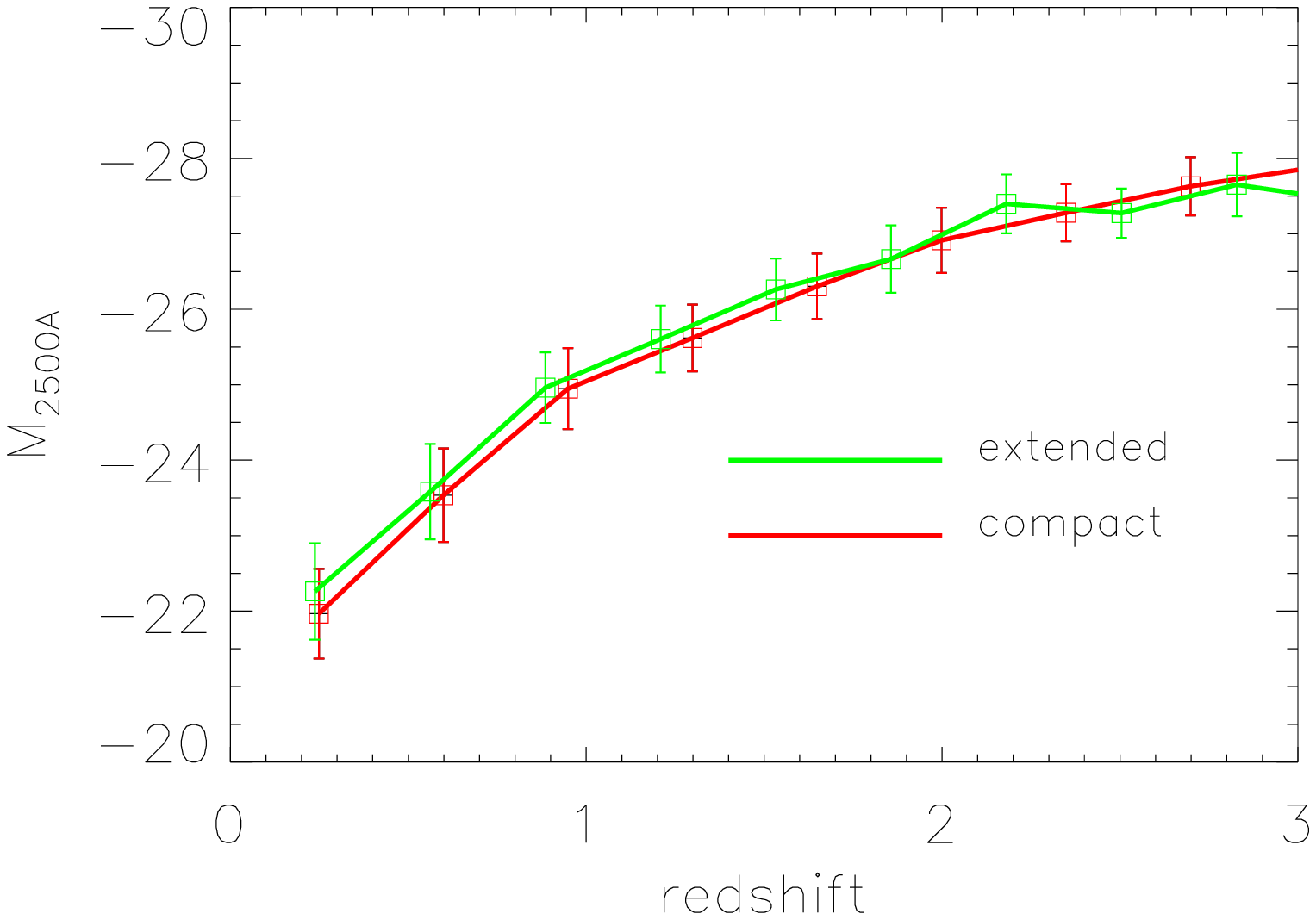}{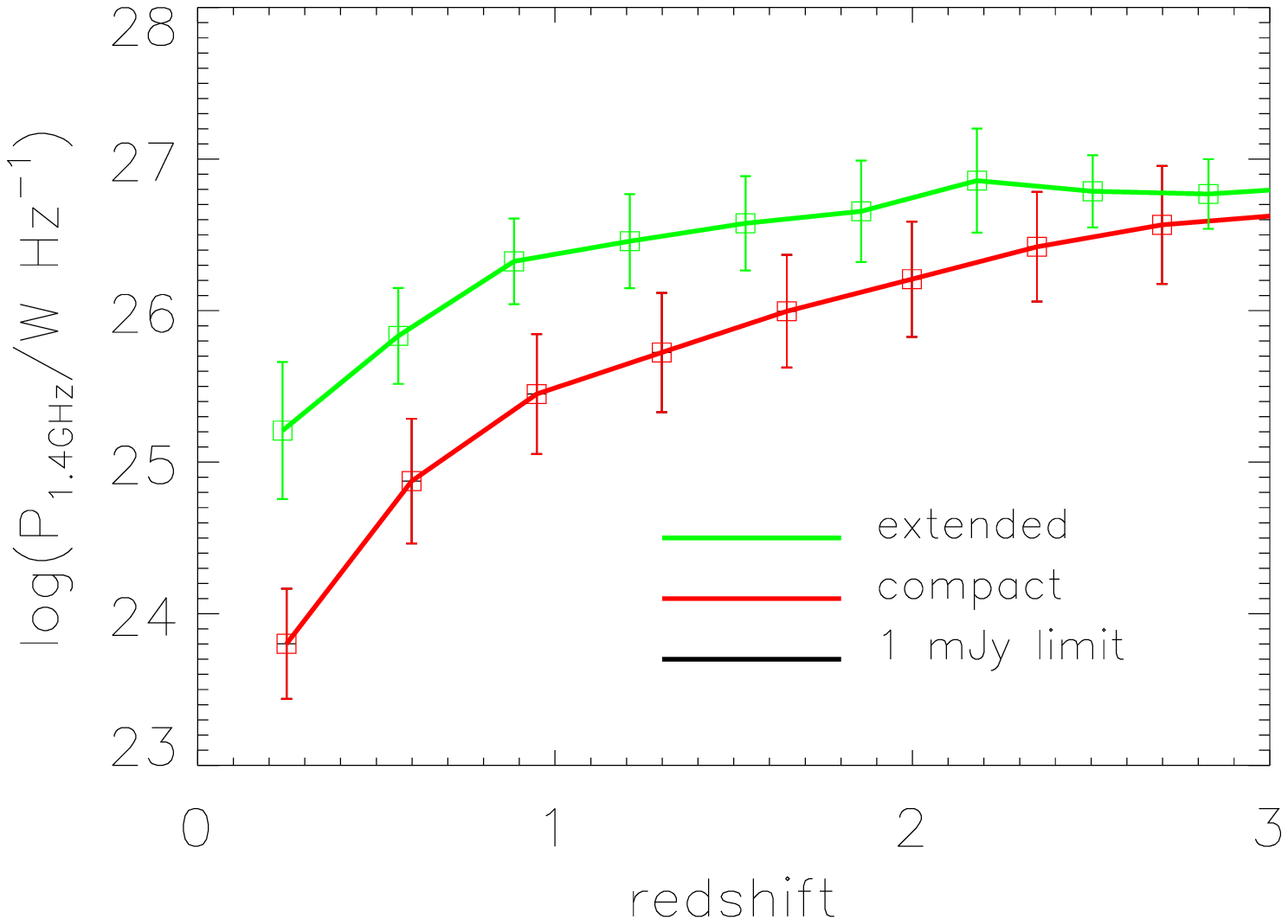} \linespread {0.3}
\caption{\label{R_O_lum} Left panel: The average absolute magnitude
at 2500A versus redshift. The green line represents the extended
sources, and the red one the compact sources. Right panel: The
average radio power of extended (green line) sources and compact
(red line) sources at 1.4GHz versus redshift. The thin dashed line
indicates the FIRST detection limit of 1mJy. Error bars are shown at
1$\sigma$ scatter around the average.}
\end{figure}

Since most quasars have redshift of $z\sim 2$, we calculate the
$k$-corrected UV flux at $2500\mathring{A}$ at the rest frame of
quasars by interpolating or extrapolating the SDSS five apparent
magnitudes using a $Spline$ function. Extrapolation is required for
only a small number of quasars at low redshift ($z<0.5$). The radio
loudness is defined as the flux ratio of $k$-corrected radio flux at
20~cm and the UV flux at  $2500~\mathring{A}$. An average of radio
spectral index $\alpha_r=-0.5$ for quasars is assumed. We plot the
average UV absolute magnitudes versus redshift in left panel of the
Fig.\ref{R_O_lum}. We find that the average UV absolute magnitudes
of extended and compact sources are similar, as they are for
resolved and unresolved sources.

\begin{figure}
\plottwo{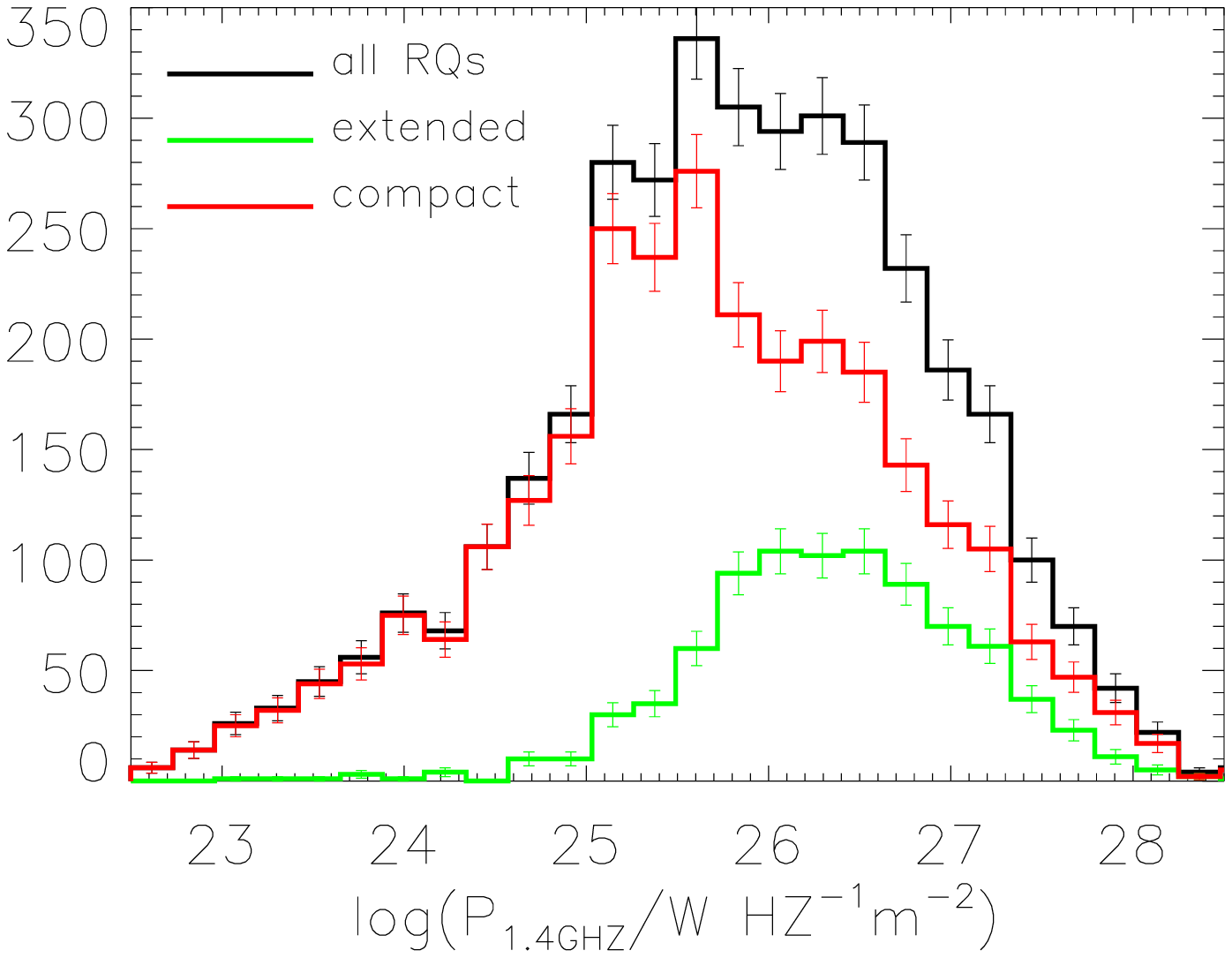}{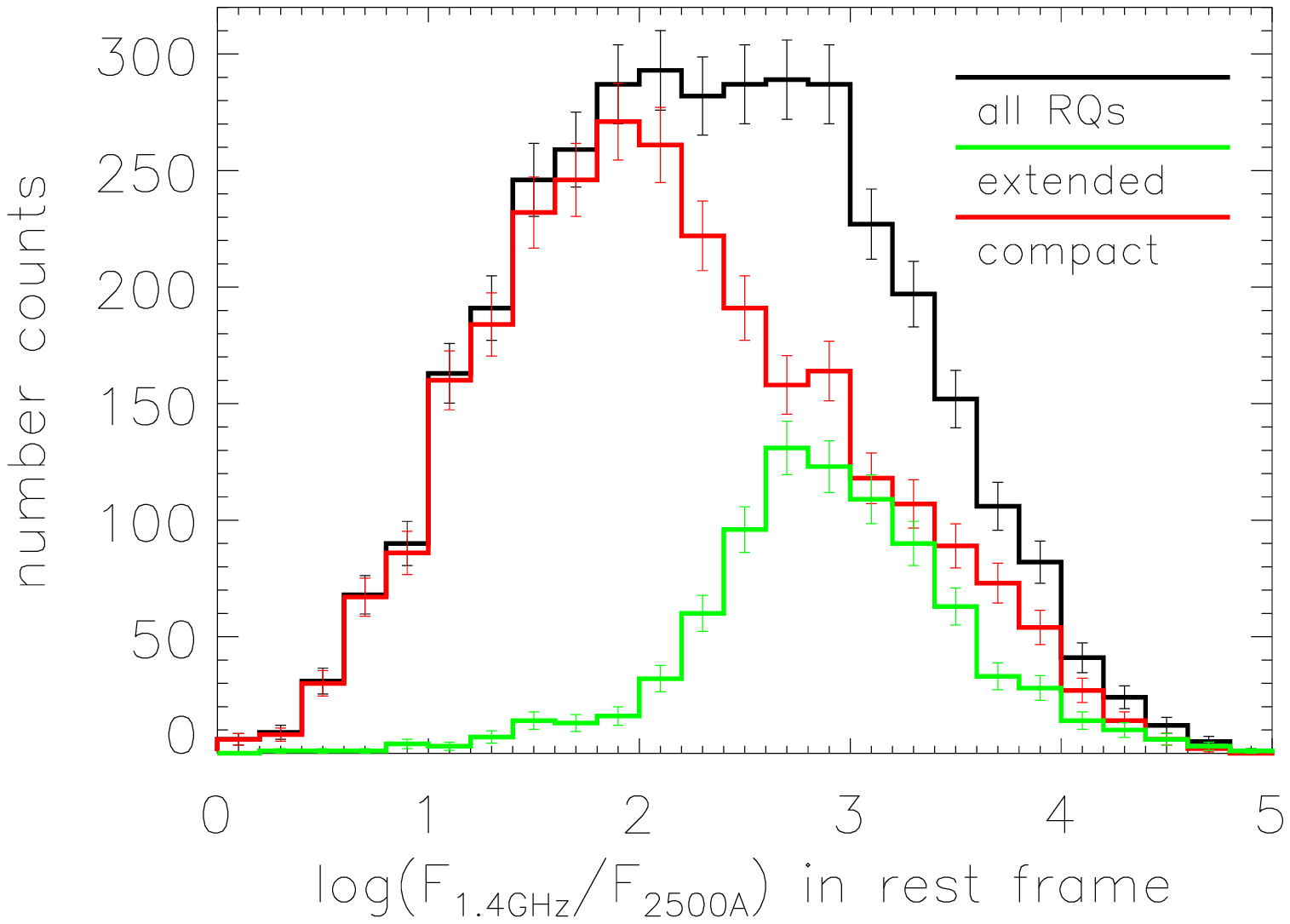} \linespread {0.3}
\caption{\label{P-R_distribution} The radio luminosity (left panel)
and the radio loudness ($f_{1.4GHz}/ f_{2500A}$ in rest frame, right
panel) distributions. The green line represents for extended radio
quasars, and the red line for compact radio quasars, and the black
line for all radio quasars.}
\end{figure}

Next, we calculated the $k$-corrected radio power at 20 cm as
$P_{radio}=4 \pi D^{2}_{L} f_{int}/ (1+z)^{1+\alpha _{r}}$, where
the radio spectral index $\alpha_{r}$ is assume to be $-0.5$ for all
objects. The radio power and the radio loudness distribution is
shown in Fig.\ref{P-R_distribution}, and the radio power as a
function $z$ is shown in right panel of Fig.\ref{R_O_lum}.
Evidently, extended radio quasars are much more powerful in the
radio than compact radio quasars despite their similar optical
luminosity.

The difference in the radio power between the extended and compact
sources decreases with increasing redshift, a factor of more than 10
at redshifts less than 0.5 to a factor $\simeq$2 or so at redshift
larger than 2 (Fig.\ref{R_O_lum}). This might be caused by a
combination of the survey limit, with which only powerful radio
sources can be detected at high $z$, and an increase in the
detecting limit of intrinsic brightness for the extended lobes at
high redshift, i.e., only very bright lobes are detected, and thus
only very powerful FR II sources. At redshift less than 0.5, the
radio power of compact radio quasars is close to the border of
FR-I/FR-II division.

It was proposed that strong radio emission from extended quasars may
be enhanced by interaction of powerful radio jets with the
interstellar medium (Bridle et al. 1995; Wills \& Brotherton 1995).
As such the difference in the radio power of core-dominated and
extended quasars is due to their different environment, rather than
their different central engine. To check this point, we compare the
core radio power for those quasars with detected cores, and found
that extended quasars have more powerful cores. Therefore,
 our result does not support this interpretation.

\begin{figure}[Ht]
\plottwo{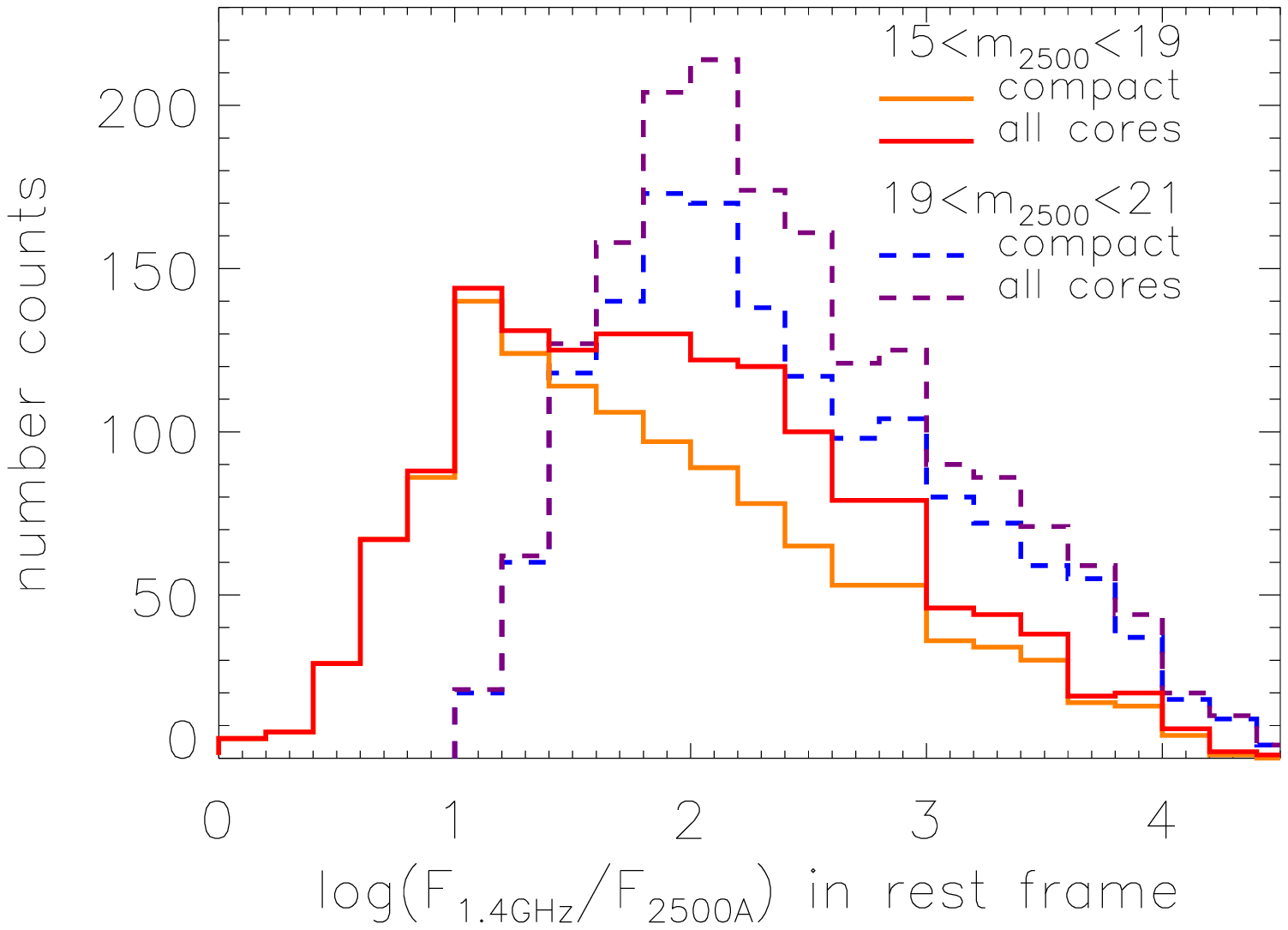}{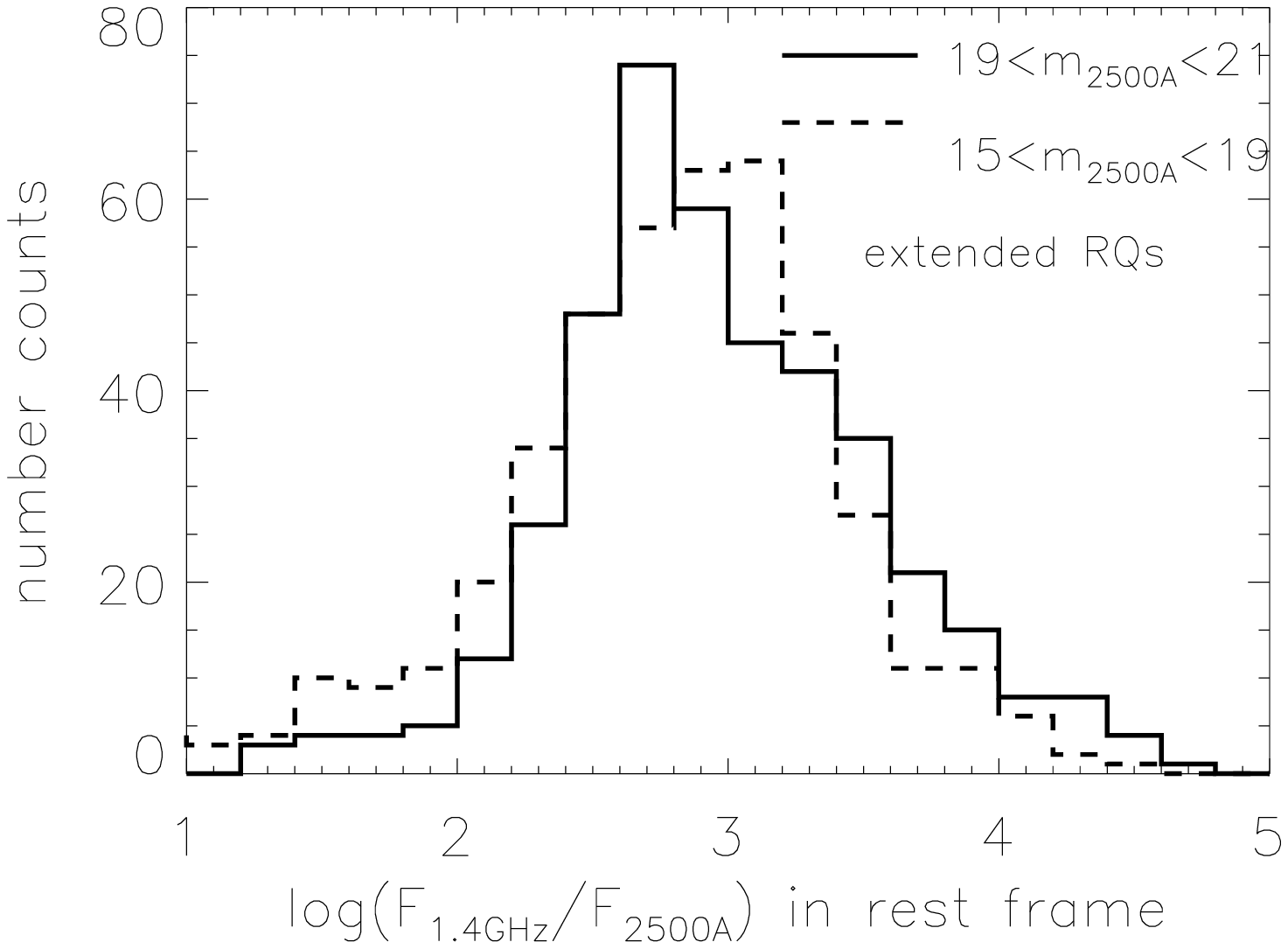} \linespread {0.3}
\caption{\label{cond_prob} The left panel: The conditional radio
loudness distribution for compact sources and for all cores
(component within 2 arcesec offset from quasar. the orange line and
red line represent compact sources and cores with $15<m_{2500}<19$,
the blue and purple line represent compact sources and cores with
$19<m_{2500}<21$. The right panel: conditional distribution of
extended sources with $15<m_{2500}<19$ (dashed line) and
$19<m_{2500}<21$ (solid line).}
\end{figure}

The radio and optical flux limits introduce another selection effect
on the radio loudness distribution of quasars. At a given optical
magnitude limit, only quasars with their radio loudness above
certain limit can be detected in the FIRST survey due its flux
limit, i.e., the sample is complete above certain radio loudness.
Using a similar strip in the log$(RL)-i$ plane as Ivezic et al.
(2004), we estimate the conditional radio loudness distribution
under different photometric limit. As shown in Fig.\ref{cond_prob},
the distribution of log$(RL)$ for compact sources peak at $\sim 2$
for $19<i<21$, consistent with Ivezic et al. (2004). Adding the core
flux of extended sources, the radio loud peak becomes more
significant with the amplitude of the dip at log$(RL)\sim2$
increasing from 24\% to 37\% due to their contribution to large
radio loudness portion. As shown in Fig.\ref{cond_prob}, the
distribution for extended quasars keeps to peak at log$(RL)\gtrsim
2.7$ at different i-magnitude bins, while the compact sources shown
peak at log$(RL)\lesssim 2.3$. Furthermore, the compact sources
distribution peaks at lower radio loudness when i decreased.

\begin{figure}[Ht]
\plottwo{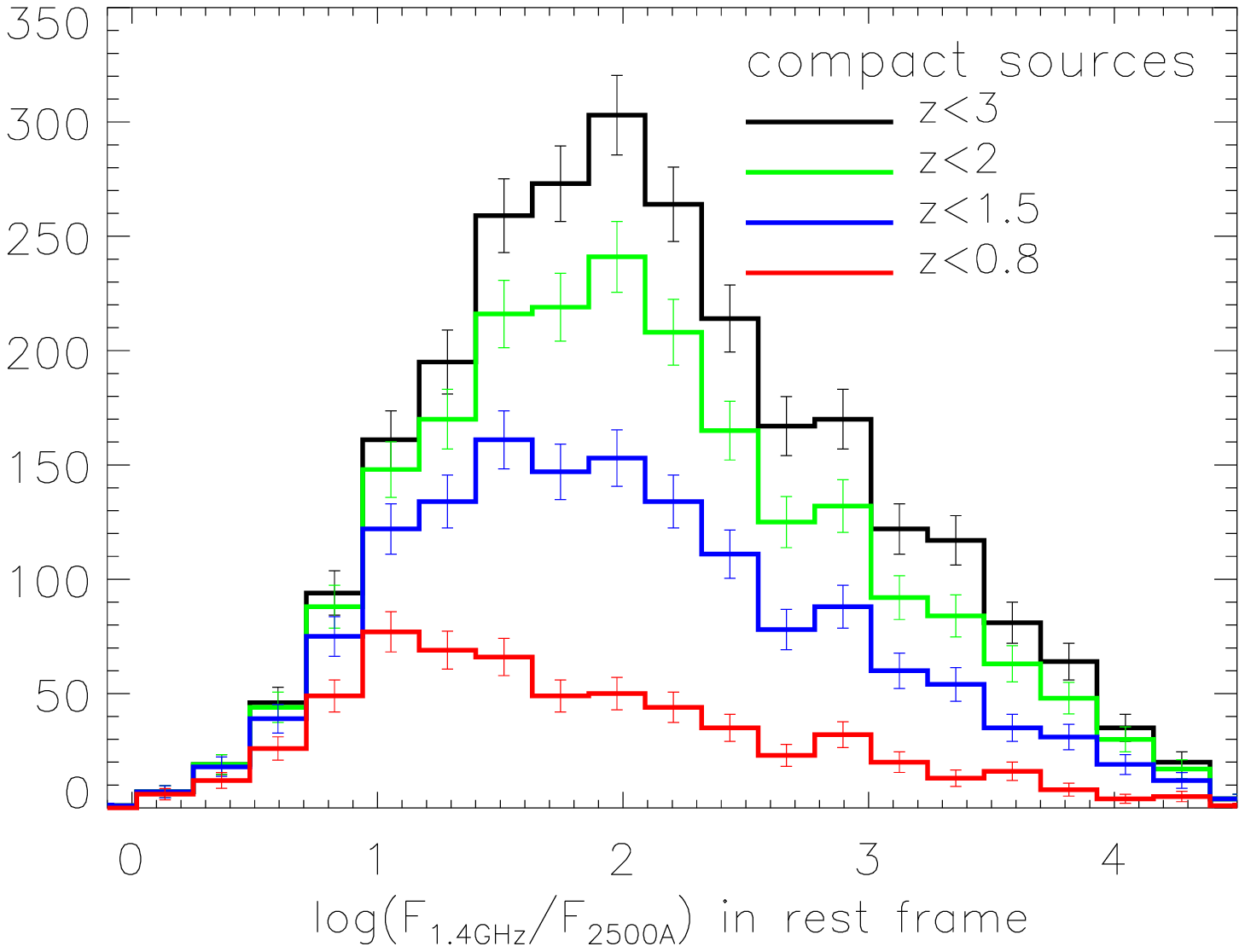}{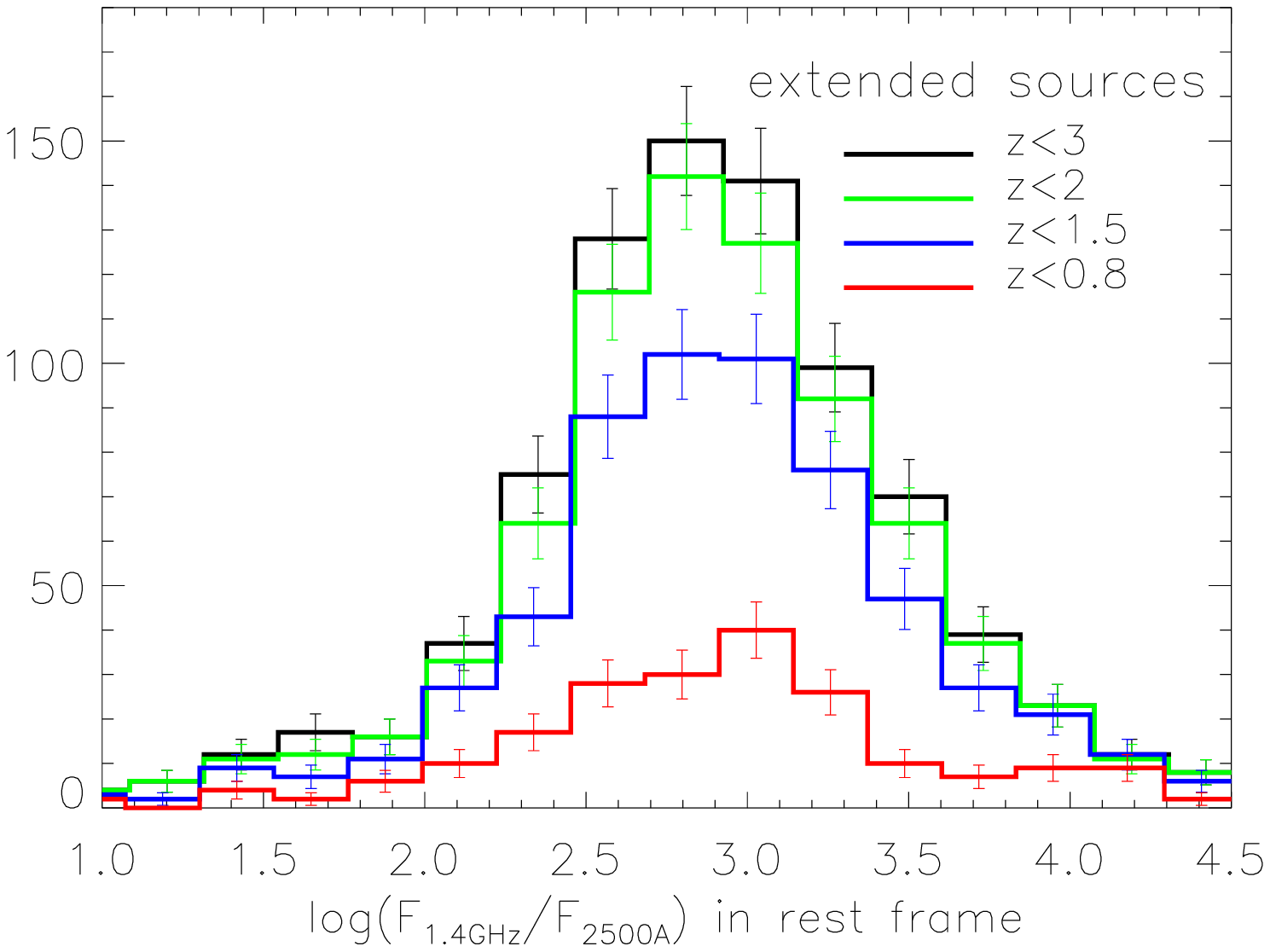} \linespread {0.3}
\caption{\label{e-c-rl2500-z} The radio loudness distribution of the
compact (left panel) and extended (right panel)
 sources below certain redshifts, the descending lines for $z<3,z<2,z<1.5,$ and $z<0.8$
, respectively.}
\end{figure}

By plotting the radio loudness under different redshift, we find
that the peak of the compact radio quasars moves to small RL as
redshift decreases, while the distribution of the extended radio
quasars remains the same (peaked at $\sim 3$) for all redshift (see
Fig.\ref{e-c-rl2500-z}) This radio loudness distribution of the
extended (also more radio loud) radio quasars is consistent with
Cirasuolo et al. (2003b), who modeled the radio loudness
distribution with double-Gaussian function by fitting the FIRST
selected 2dF quasars, and found that the intrinsic radio loudness of
radio quasars peaks at log$(RL)=2.7\pm 0.2$ and log$(RL)=-0.5\pm
0.3$.

\section{Conclusion and Discussion}

We have constructed a relatively unbiased large radio quasar sample
using the SDSS quasar catalog (Schneider et al. 2005), and the FIRST
catalog and images. Apart from positional coincidence of radio
sources within $2^{''}$ of quasars, we also identify the radio
counterpart of quasars with complex radio morphologies such as lobe
-dominated quasars by visual inspection of their radio images. We
find that using the positional coincidence alone will miss $\sim
8\%$ radio counterpart that do not show radio core at FIRST flux
limit of 1 mJy, and under-estimates the radio flux by a factor of
more than two in another $\sim 9\%$ objects. By comparing the radio
flux from FIRST survey with that from NVSS, we found that lobes in
weak radio sources tend to be missed in this sample. So these
numbers are only lower limits.

Quasars with extended radio emission show both larger radio powers
and radio loudness, and appear somewhat bluer than radio compact
quasars despite their indistinguishable optical luminosity. As such,
the radio extended quasars account for nearly one third of radio
loud quasars at log$(RL)>2.2$. Naturally, including the extended
emission and weak core radio sources increases the fraction of the
radio loud objects and the significance of radio loud peak in the
distribution of Hough \& Readhead radio loudness.

Our results in the first glance are not consistent with the simple
unification scheme in which radio compact quasars are extended ones
viewed along radio jet, for which the relativistic beaming enhances
the core radio emission and as such the total radio power (Wills \&
Browne 1986; Hough \& Readhead 1988; Barthel 1989; Falcke et al.
2004) when the projection effect would make the apparent size
smaller. Within such a scheme, the unresolved core is enhanced
because of the beaming effect. That the lobe-dominated quasars are
more luminous in radio seems to contradict this model. However,
there are at least two selection effects that make the average radio
power in the core-dominated sources smaller.

\begin{figure}
\plottwo{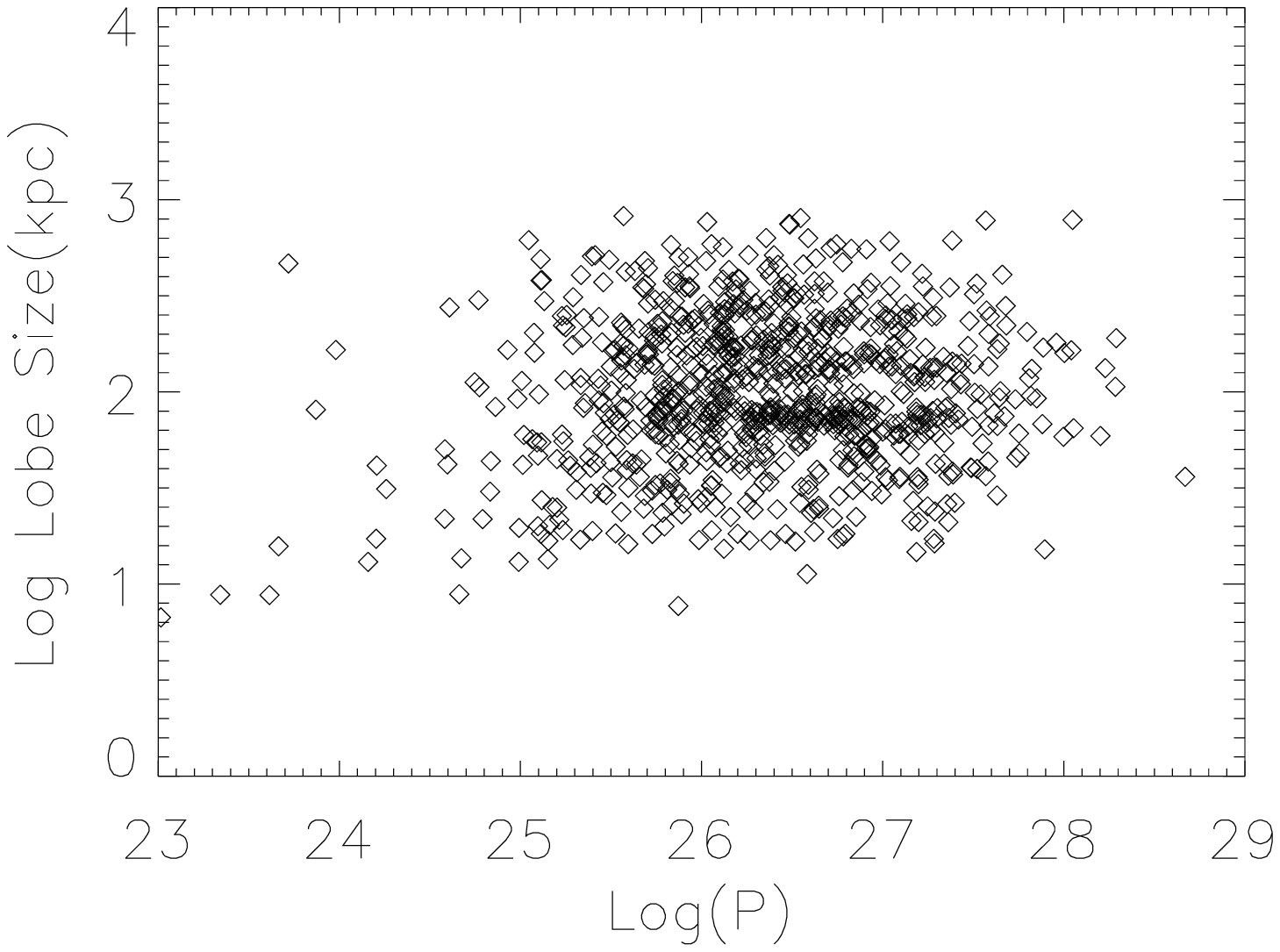}{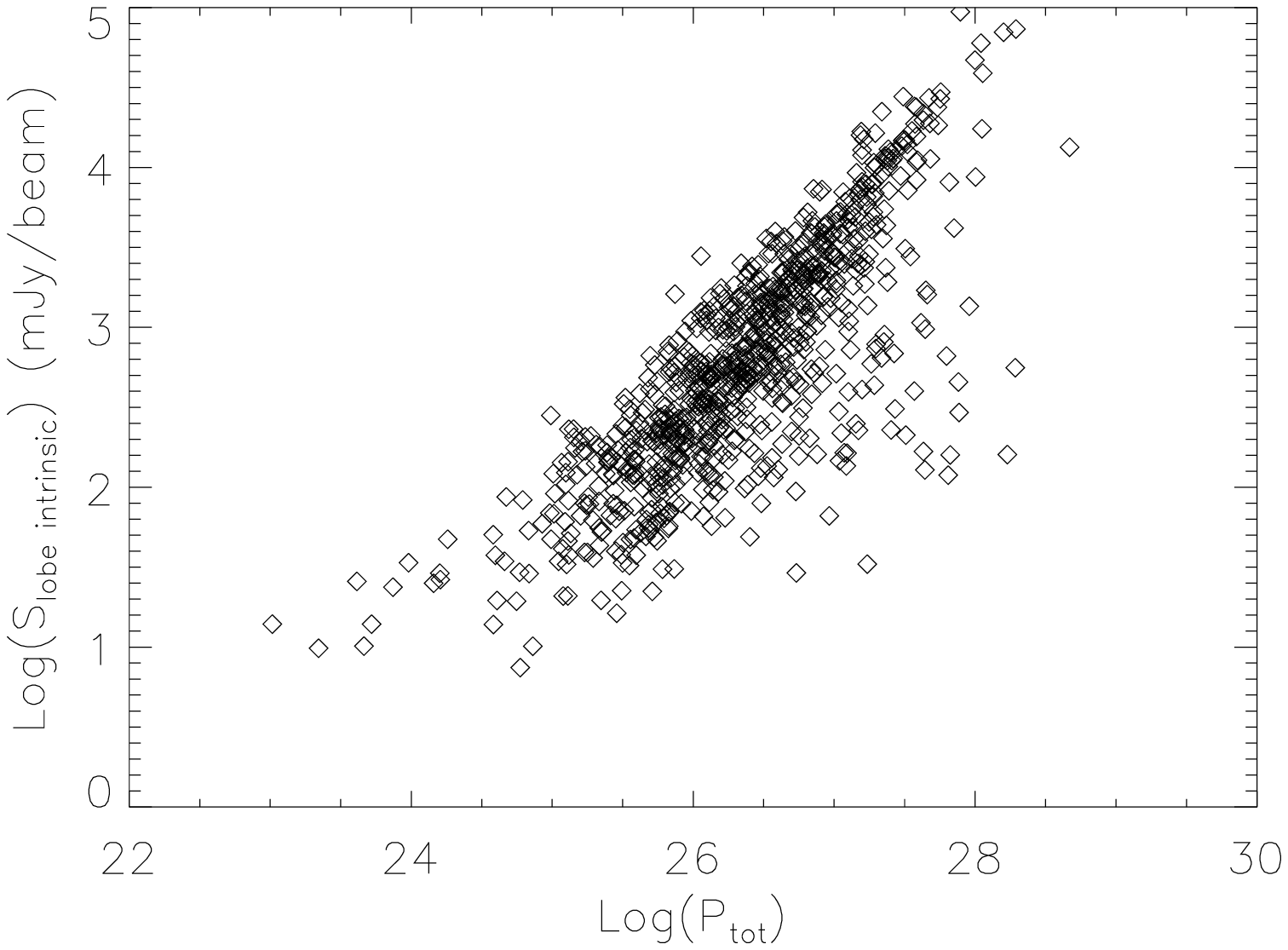}
\caption{\label{lobe-distrib}
Left panel: The distance of the furthest radio lobe to the quasar
versus the total radio power for extended radio quasars. There is no
apparent correlation. Right panel: The intrinsic peak brightness of
resolved lobes versus the total radio power (defined as $P_{tot}$)
for extended radio quasars.}
\end{figure}

First, as we showed in the last section that the peak brightness of radio
lobe component is correlated with radio power, and the FIRST will not be able
to detect the lobe component in less radio power sources, especially at
high redshift (see also Fig.\ref {lobe-distrib}). Second, if most of
core-dominated quasars are intrinsically radio weaker (Wang et al. 2006)
and if the radio luminosity function of them is steep, their average
apparent luminosity can be lower even if the radio power had been boosted.

The relative number density of the extended and compact sources also
suggests that majority of compact sources are either of beamed,
intrinsically much weak radio quasars or intrinsically compact radio
sources, such as CSS (Compact Steep Spectrum Objects) or GPS
(GigaHertz Peaked Sources). With typical Lorentz factor of 10-15 for
jets in FR-II radio quasars, the boosted emission can be viewed in
only relatively small fraction of solid angle between the line of
sight and the jet, $\theta <7^{\circ} $, whereas at other angles the
core is weakened due to Doppler effect. If the power of the
un-beamed radio cores is near 0.005 of that of the lobes, as
determined for 3CR sources (Urry \& Padovani 1995), and if the
intrinsic radio loudness distribution following Cirasuolo et al.
(2003b) , and with the luminosity distribution as DR3 quasar sample,
we can estimate that $\sim 62\%$ beamed intrinsic radio quiet
sources could be detected by FIRST. Since GPS and CSS are all
powerful radio sources, it is likely most of these compact quasars
are beamed radio intermediate quasars as proposed by Falcke et al.
(1996)

Our results suggest that lobe-dominated sources are not particularly reddened,
in agreement with the finding by  de Vries et al. (2005). This is valid even
for quasars without detectable radio core at flux down to the FIRST limit. The line of
sight does not intercept the dusty torus in those lobe-dominated quasars.
However, Backer et al. (1997) found that most lobe-dominated quasars in the Molonglo quasars are reddened by $A_V\simeq 2-4$, and CSSs are most
reddened. It should be pointed out that quasars reddened by this amount cannot
 be found in the `color' selected sample, in particularly at high redshift due
to strong attenuation in ultraviolet. The slightly bluer color for
lobe-dominated quasars in this sample could be due to inclusion of
CSS-like objects in the core-dominant objects or due to a selection
effect by which reddened "extended" radio quasars are lost.

If Baker (1997) is correct, we might miss a large number of heavily
reddened lobe-dominated quasars. Although most such quasars are likely
below the magnitude limit of spectroscopic quasar sample, in principle
the FIRST selected sample is able to detect some of such reddened quasars,
particularly in the low redshift if a weak core is present. We look at the
spectroscopic sources that selected as FIRST sources only, and find
that the FIRST-only selected spectroscopic sources are indeed much
redder. But the fraction of quasars with extended lobes in FIRST-only sources
is very low($\sim 0.2\%$) probably due to large extinction. Thus it is not
conclusive whether a large number of such reddened lobe-dominated quasars
do exist.

\acknowledgements We thank the anonymous referee for the constructive comments.
This work is supported by Chinese Natural Science Foundation through
CNSF-10233030 and CNSF-10573015.

\begin{thebibliography}{}

\bibitem[Aars et al.(2005)]{2005AJ....130...23A} Aars, C.~E., Hough, D.~H.,
Yu, L.~H., Linick, J.~P., Beyer, P.~J., Vermeulen, R.~C., \& Readhead,
A.~C.~S.\ 2005, \aj, 130, 23

\bibitem[Abazajian et al.(2005)]{2005AJ....129.1755A} Abazajian, K., et
al.\ 2005, \aj, 129, 1755

\bibitem[Barthel(1989)]{1989ApJ...336..606B} Barthel, P.~D.\ 1989, \apj,
336, 606

\bibitem[Becker et al.(1995)]{1995ApJ...450..559B} Becker, R.~H., White,
R.~L., \& Helfand, D.~J.\ 1995, \apj, 450, 559

\bibitem[Bennett(1962)]{1962MmRAS..68..163B} Bennett, A.~S.\ 1962, \memras,
68, 163

 \bibitem[Best et al.(2005)]{2005MNRAS.362....9B} Best, P.~N., Kauffmann,
G., Heckman, T.~M., \& Ivezi{\'c}, {\v Z}.\ 2005, \mnras, 362, 9

\bibitem[Backer et al.(1997)]{1997PASP..109...61B} Backer, D.~C., Dexter,
M.~R., Zepka, A., Ng, D., Werthimer, D.~J., Ray, P.~S., \& Foster, R.~S.\
1997, \pasp, 109, 61


\bibitem[Boroson(2002)]{2002ApJ...565...78B} Boroson, T.~A.\ 2002, \apj,
565, 81


\bibitem[Boyle et al.(2001)]{2001defi.conf..282B} Boyle, B.~J., Croom,
S.~M., Smith, R.~J., Shanks, T., Outram, P.~J., Hoyle, F., Miller, L., \&
Loaring, N.~S.\ 2001, Deep Fields, 282


\bibitem[Bridle et al.(1994)]{1994AJ....108..766B} Bridle, A.~H., Hough,
D.~H., Lonsdale, C.~J., Burns, J.~O., \& Laing, R.~A.\ 1994, \aj, 108, 766

\bibitem[Brinkmann et al.(2000)]{2000yCat..33560445B} Brinkmann, W.,
Laurent-Muehleisen, S.~A., Voges, W., Siebert, J., Becker, R.~H.,
Brotherton, M.~S., White, R.~L., \& Gregg, M.~D.\ 2000, VizieR Online Data
Catalog, 335, 60445

\bibitem[Cirasuolo et al.(2003)]{2003MNRAS.341..993C} Cirasuolo, M.,
Magliocchetti, M., Celotti, A., \& Danese, L.\ 2003, \mnras, 341, 993


\bibitem[Cirasuolo et al.(2003)]{2003MNRAS.346..447C} Cirasuolo, M.,
Celotti, A., Magliocchetti, M., \& Danese, L.\ 2003, \mnras, 346, 447

\bibitem[Condon et al.(1998)]{1998AJ....115.1693C} Condon, J.~J., Cotton,
W.~D., Greisen, E.~W., Yin, Q.~F., Perley, R.~A., Taylor, G.~B., \&
Broderick, J.~J.\ 1998, \aj, 115, 1693

\bibitem[Colla et al.(1972)]{1972A&AS....7....1C} Colla, G., et al.\ 1972,
\aaps, 7, 1

\bibitem[Cress et al.(1995)]{1995AAS...187.5402C} Cress, C.~M., Helfand,
D.~J., Becker, R.~H., \& White, R.~L.\ 1995, Bulletin of the American
Astronomical Society, 27, 1364

\bibitem[Cress et al.(1996)]{1996ApJ...473....7C} Cress, C.~M., Helfand,
D.~J., Becker, R.~H., Gregg, M.~D., \& White, R.~L.\ 1996, \apj, 473, 7

\bibitem[de Vries et al.(2006)]{2006AJ....131..666D} de Vries, W.~H.,
Becker, R.~H., \& White, R.~L.\ 2006, \aj, 131, 666

 \bibitem[Eisenstein et al.(2001)]{2001AJ....122.2267E} Eisenstein, D.~J.,
et al.\ 2001, \aj, 122, 2267

\bibitem[Falcke et al.(1995)]{1995A&A...298..395F} Falcke, H.,
Gopal-Krishna, \& Biermann, P.~L.\ 1995, \aap, 298, 395

\bibitem[Falcke et al.(1996)]{1996ApJ...471..106F} Falcke, H., Sherwood,
W., \& Patnaik, A.~R.\ 1996, \apj, 471, 106


\bibitem[Falcke et al.(2004)]{2004A&A...414..895F} Falcke, H., K{\"o}rding,
E., \& Markoff, S.\ 2004, \aap, 414, 895

\bibitem[Fanaroff \& Riley(1974)]{1974MNRAS.167P..31F} Fanaroff, B.~L., \&
Riley, J.~M.\ 1974, \mnras, 167, 31P

\bibitem[Fanti et al.(1974)]{1974A&AS...18..147F} Fanti, C., Fanti, R.,
Ficarra, A., \& Padrielli, L.\ 1974, \aaps, 18, 147

\bibitem[Gregg et al.(1996)]{1996AJ....112..407G} Gregg, M.~D., Becker,
R.~H., White, R.~L., Helfand, D.~J., McMahon, R.~G., \& Hook, I.~M.\ 1996,
\aj, 112, 407

\bibitem[Hardcastle et al.(1998)]{1998MNRAS.296..445H} Hardcastle, M.~J.,
Alexander, P., Pooley, G.~G., \& Riley, J.~M.\ 1998, \mnras, 296, 445

\bibitem[Hewett et al.(2001)]{2001AJ....122..518H} Hewett, P.~C., Foltz,
C.~B., \& Chaffee, F.~H.\ 2001, \aj, 122, 518

\bibitem[Hoekstra et al.(1997)]{1997A&A...319..757H} Hoekstra, H., Barthel,
P.~D., \& Hes, R.\ 1997, \aap, 319, 757

\bibitem[Hough \& Readhead(1988)]{1988IAUS..129...99H} Hough, D.~H., \&
Readhead, A.~C.~S.\ 1988, IAU Symp.~129: The Impact of VLBI on Astrophysics
and Geophysics, 129, 99

\bibitem[Ivezi{\'c} et al.(2002)]{2002AJ....124.2364I} Ivezi{\'c}, {\v Z}.,
et al.\ 2002, \aj, 124, 2364

\bibitem[Ivezi{\'c} et al.(2004)]{2004ASPC..311..347I} Ivezi{\'c}, Z., et
al.\ 2004, ASP Conf.~Ser.~311: AGN Physics with the Sloan Digital Sky
Survey, 311, 347

\bibitem[Jackson \& Wall(1998)]{1998ocnr.conf..203J} Jackson, C.~A., \&
Wall, J.~V.\ 1998, ASSL Vol.~226: Observational Cosmology with the New
Radio Surveys, 203

\bibitem[Jackson \& Wall(1999)]{1999MNRAS.304..160J} Jackson, C.~A., \&
Wall, J.~V.\ 1999, \mnras, 304, 160

\bibitem[Jackson(1999)]{1999PASA...16..124J} Jackson, C.~A.\ 1999,
Publications of the Astronomical Society of Australia, 16, 124

\bibitem[Kaiser(2000)]{2000A&A...362..447K} Kaiser, C.~R.\ 2000, \aap, 362,
447

\bibitem[Kellermann et al.(1989)]{1989AJ.....98.1195K} Kellermann, K.~I.,
Sramek, R., Schmidt, M., Shaffer, D.~B., \& Green, R.\ 1989, \aj, 98, 1195

\bibitem[Knapp et al.(2002)]{2002AAS...201.4804K} Knapp, G.~R., et al.\
2002, Bulletin of the American Astronomical Society, 34, 1180

\bibitem[Krolik(1999)]{1999agnc.book.....K} Krolik, J.~H.\ 1999, Active
galactic nuclei : from the central black hole to the galactic environment
.Princeton, N.~J.~: Princeton University Press


 \bibitem[Lacy \& Ridgway(2001)]{2001AAS...19914109L} Lacy, M., \& Ridgway,
S.~E.\ 2001, Bulletin of the American Astronomical Society, 33, 1520

\bibitem[Large et al.(1981)]{1981MNRAS.194..693L} Large, M.~I., Mills,
B.~Y., Little, A.~G., Crawford, D.~F., \& Sutton, J.~M.\ 1981, \mnras, 194,
693
\bibitem[McMahon et al.(2002)]{2002ApJS..143....1M} McMahon, R.~G., White,
R.~L., Helfand, D.~J., \& Becker, R.~H.\ 2002, \apjs, 143, 1

\bibitem[Maddox(1998)]{1998ASPC..146..198M} Maddox, S.\ 1998, ASP
Conf.~Ser.~146: The Young Universe: Galaxy Formation and Evolution at
Intermediate and High Redshift, 146, 198

\bibitem[Magliocchetti \& Maddox(2002)]{2002MNRAS.330..241M} Magliocchetti,
M., \& Maddox, S.~J.\ 2002, \mnras, 330, 241

\bibitem[McMahon et al.(2002)]{2002ApJS..143....1M} McMahon, R.~G., White,
R.~L., Helfand, D.~J., \& Becker, R.~H.\ 2002, \apjs, 143, 1

\bibitem[Meier(2001)]{2001ApJ...548L...9M} Meier, D.~L.\ 2001, \apjl, 548,L9

\bibitem[Miller et al.(1990)]{1990MNRAS.244..207M} Miller, L., Peacock,                                                                         J.~A., \& Mead, A.~R.~G.\ 1990, \mnras, 244, 207
\bibitem[Padovani et al.(2003)]{2003ASPC..299...63P} Padovani, P.,

Costamante, L., Ghisellini, G., Giommi, P., \& Perlman, E.\ 2003, ASP
Conf.~Ser.~299: High Energy Blazar Astronomy, 299, 63

\bibitem[Readhead et al.(1988)]{1988IAUS..129...65R} Readhead, A.~C.~S.,
Pearson, T.~J., \& Barthel, P.~D.\ 1988, IAU Symp.~129: The Impact of VLBI
on Astrophysics and Geophysics, 129, 65

\bibitem[Richards et al.(2002)]{2002AJ....123.2945R} Richards, G.~T., et
al.\ 2002, \aj, 123, 2945

\bibitem[Richards et al.(2006)]{2006AJ....131.2766R} Richards, G.~T., et
al.\ 2006, \aj, 131, 2766

\bibitem[Schneider et al.(2005)]{2005AJ....130..367S} Schneider, D.~P., et
al.\ 2005, \aj, 130, 367

\bibitem[Shastri et al.(1993)]{1993ApJ...410...29S} Shastri, P., Wilkes,
B.~J., Elvis, M., \& McDowell, J.\ 1993, \apj, 410, 29

\bibitem[Smith et al.(1976)]{1976PASP...88..621S} Smith, H.~E., Smith,
E.~O., \& Spinrad, H.\ 1976, \pasp, 88, 621

\bibitem[Stoughton et al.(2002)]{2002AJ....123..485S} Stoughton, C., et
al.\ 2002, \aj, 123, 485

\bibitem[Strauss et al.(2002)]{2002AJ....124.1810S} Strauss, M.~A., et al.\
2002, \aj, 124, 1810

\bibitem[Urry \& Padovani(1995)]{1995PASP..107..803U} Urry, C.~M., \&
Padovani, P.\ 1995, \pasp, 107, 803

\bibitem[Wang et al. (2006)]{2006APJ accepted} Wang, T.~G., Zhou, H.~Y., Wang, J.~X., Ru,Y., Ru.Y.,\& Lu,Y.~J. 2006. APJ accepted

\bibitem[White(1999)]{1999hst..prop.4716W} White, R.\ 1999, HST Proposal,
4716

\bibitem[White et al.(2000)]{2000ApJS..126..133W} White, R.~L., et al.\
2000, \apjs, 126, 133

\bibitem[Wills \& Browne(1986)]{1986ApJ...302...56W} Wills, B.~J., \&
Browne, I.~W.~A.\ 1986, \apj, 302, 56

\bibitem[Wills \& Brotherton(1995)]{1995ApJ...448L..81W} Wills, B.~J., \&
Brotherton, M.~S.\ 1995, \apjl, 448, L81

\bibitem[Wilson \& Colbert(1995)]{1995ApJ...438...62W} Wilson, A.~S., \&
Colbert, E.~J.~M.\ 1995, \apj, 438, 62


\bibitem[York et al.(2000)]{2000AJ....120.1579Y} York, D.~G., et al.\ 2000,
\aj, 120, 1579


\end{thebibliography}
\end{document}